\documentclass
{aastex63}
\usepackage{amsmath,amsthm,amssymb}
\usepackage{graphicx}
\usepackage{bm}
\usepackage{slashed}
\usepackage{textcomp}
\usepackage{threeparttable}
\usepackage{booktabs}
\usepackage{latexsym}
\usepackage{mciteplus}
\usepackage{amsmath}
\usepackage{color}
\usepackage{hyperref}
\renewcommand{\TPTtagStyle}%
{\normalsize\textit}
\usepackage[normalem]{ulem} 
\DeclareFontFamily{T1}{calligra}{}

\DeclareRobustCommand{\sr}{%
  \mspace{-2mu}%
  \text{\usefont{T1}{calligra}{m}{n}r\/}%
  \mspace{2mu}%
}

\begin{document}

\title{Chiral Radiation Transport Theory of Neutrinos}
\date{\today}
\author{Naoki Yamamoto and Di-Lun Yang}
\affiliation{Department of Physics,  Keio University, Yokohama 223-8522, Japan }
\begin{abstract}
We construct the chiral radiation transport equation for left-handed neutrinos in the context of radiation hydrodynamics for core-collapse supernovae. Based on the chiral kinetic theory incorporating quantum corrections due to the chirality of fermions, we derive a general relativistic form of the chiral transfer equation with collisions. We show that such quantum corrections explicitly break the spherical symmetry and axisymmetry of the system. In the inertial frame, in particular, we find that the so-called side jump leads to quantum corrections in the collisions between neutrinos and matter. We also derive analytic forms of such corrections in the emission and absorption rates for the neutrino absorption process. These corrections result in the generation of kinetic helicity and cross helicity of matter, which should then modify the subsequent evolution of matter. This theoretical framework can be applied to investigate the impacts of the chirality of neutrinos on the evolution of core-collapse supernovae.   	
\end{abstract}

\section{Introduction}
Understanding the mechanism of core-collapse supernova explosions is one of the unsolved problems in astrophysics. When a massive star experiences collapse of the core, most of the gravitational binding energy is released in the form of neutrinos. For this reason, proper treatment of neutrino transport physics is required to account for the core-collapse supernova explosions. Since neutrinos are mostly out of equilibrium and must be treated as radiation away from the dense core of supernovae, the theoretical formulations and numerical simulations for neutrino transport are based on the Boltzmann equation, or more precisely, the Einstein-Vlasov equation under certain approximations \citep{Castor_1972ApJ...178..779C,Bruenn:1985en}; see also \citet{O_Connor_2018,O_Connor_2018_2,Summa_2018,Richers_2017,10.1093/mnras/sty2585,Kotake_2018,Cabezon:2018lpr} for recent reviews and comparisons between numerical simulations.%
\footnote{More recently, the effects of neutrino oscillations in the neutrino self-energy in quantum kinetic theory have also been studied \citep{Vlasenko:2013fja,Cirigliano:2014aoa,Kartavtsev:2015eva,Blaschke:2016xxt,Richers:2019grc}, which is a different kind of quantum corrections from those we would like to address in this paper.} 
However, the most fundamental property of neutrinos---left-handedness---has been neglected in the conventional theoretical formulation and simulations of radiation hydrodynamics for neutrinos.

Recently, it has been shown in \citet{Yamamoto:2015gzz} that the parity violation by the chirality of neutrinos can affect the macroscopic hydrodynamic evolution of supernovae in a qualitative manner. In fact, there has been growing recent interest in the study of chiral transport phenomena that originate from chirality of (generally charged) particles not limited to neutrinos. The most renowned examples are the currents induced by magnetic fields and vorticity, dubbed the chiral magnetic effect (CME) \citep{Vilenkin:1980fu,Nielsen:1983rb,Alekseev:1998ds,Fukushima:2008xe} and chiral vortical effect (CVE) \citep{Vilenkin:1979ui,Erdmenger:2008rm,Banerjee:2008th,Son:2009tf,Landsteiner:2011cp}, respectively. A remarkable aspect of these effects is their connection to the chiral anomaly, i.e., the quantum violation of the chiral symmetry in field theory \citep{Adler:1969gk,Bell:1969ts}. Such anomalous transport phenomena are relevant not only to neutrinos in core-collapse supernovae but also to a variety of physical systems such as hot electroweak plasmas in the early universe \citep{Joyce:1997uy,Boyarsky:2011uy,Kamada:2016eeb}, quark-gluon plasmas created in heavy ion collision experiments \citep{Kharzeev:2015znc}, dense electromagnetic plasmas in neutron stars \citep{Charbonneau:2009ax,Akamatsu:2013pjd,Ohnishi:2014uea,Kaminski:2014jda}, and emergent chiral matter near band crossing points of Weyl semimetals \citep{Nielsen:1983rb,Wan:2011udc,Burkov:2011ene,Xu:2011dn}.%
\footnote{In the context of core-collapse supernovae, there could also be prominent chirality imbalance of electrons produced by the electron capture process \citep{Ohnishi:2014uea,Dvornikov:2014uza}. Although such chiral imbalance could be compensated by elastic electron scattering with the effect of nonzero electron mass \citep{Grabowska:2014efa,Kaplan:2016drz}, the remaining imbalance may still result in sizable chiral effects \citep{Sigl:2015xva,Onishi:2020rqr}. To investigate the dynamics of the chiral matter near the dense core of supernovae in thermal equilibrium, one may resort to the chiral magnetohydrodynamics (ChMHD) as the modified magnetohydrodynamics involving the chiral anomaly \citep{Yamamoto:2015gzz,Yamamoto:2016xtu,Rogachevskii:2017uyc,Hattori:2017usa}. It has been demonstrated in \citet{Masada:2018swb} that the ChMHD simulation reveals the dominance of inverse energy cascade as opposed to of the direct energy cascade in conventional 3D neutrino radiation hydrodynamic simulations \citep[for reviews, see][]{Janka:2016fox,Radice:2017kmj}. In this paper, we will focus on the chiral effects of neutrinos.}

Nevertheless, the classical Boltzmann equation is unable to capture these chiral effects. To incorporate such quantum corrections, 
the so-called chiral kinetic theory (CKT) has been established. The pioneering construction started from a semiclassical derivation by introducing a Berry phase as the source of quantum corrections, which results in the modification on the free-streaming Boltzmann equation \citep{Son:2012wh,Stephanov:2012ki}. Alternatively, a field-theoretic  derivation known as the Wigner function approach was applied to derive CKT despite some limited conditions \citep{Son:2012zy,Chen:2012ca}. In addition, the Lorentz invariance of the CKT was revealed and the modified frame transformation on distribution functions was introduced in relation to the so-called side-jump phenomenon stemming from the spin-orbit interaction \citep{Chen:2014cla,Chen:2015gta,Hidaka:2016yjf,Yang:2018lew}. More recently, through the Wigner function approach, a generic Lorentz-covariant CKT under background electromagnetic fields with systematic inclusion of collisions was obtained \citep{Hidaka:2016yjf,Hidaka:2017auj,Hidaka:2018ekt}; see also \citet{Mueller:2017arw, Huang:2018wdl, Carignano:2018gqt, Carignano:2019zsh, Lin:2019ytz,Lin:2019fqo} for related recent developments. Furthermore, this derivation was generalized to curved spacetime in the case without collisions \citep{Liu:2018xip}. The CKT has been widely applied to investigate anomalous transport pertinent to relativistic heavy ion collisions and Weyl semimetals \citep{Gorbar:2016ygi,Kharzeev:2016sut,Huang:2017tsq,Hidaka:2018ekt,Rybalka:2018uzh,Sun:2018idn}.

Given the established framework of CKT and Wigner functions with quantum corrections, in this paper we construct the radiation transport equation for left-handed neutrinos by incorporating the effects of the chirality, which we may call the chiral radiation transport (or transfer) equation. We first derive a general relativistic form of the chiral radiation transport equation with collisions (Equations~(\ref{CKT_noBF}) and (\ref{Gamma})), which shows that the quantum corrections explicitly break the spherical symmetry and axisymmetry of the system. We then focus on the inertial frame as one of widely used coordinate systems for numerical simulations of core-collapse supernovae. In this case, we find that, although the free-streaming part remains unchanged from the conventional transport equation, the so-called side-jump effects lead to quantum corrections between neutrinos and matter (Equations~(\ref{CKT_inertial_con}) and (\ref{Gamma_inertial})). As a demonstration, we analytically derive the quantum corrections involving the fluid vorticity and magnetic fields in the emission and absorption rates for the neutrino absorption process (Equations~(\ref{Gamma_decomp})--(\ref{Gamma_omega})). In addition, we also show that the side-jump effects modify the particle-number current and energy-momentum tensor of neutrinos through the Wigner functions (Equations~(\ref{J_and_T}) and (\ref{Lf_inetrial})) and that such quantum corrections affect the energy-momentum transfer between neutrinos and matter (Equations~(\ref{cons_Tmunu}) and (\ref{C_check}); see also Equation~(\ref{cons_Tmunu2})). 

The paper is organized as follows: In Section \ref{sec_Boltzmann}, we briefly review the derivation of 3D transfer equations from the Einstein-Vlasov equation mainly in the inertial frame. In Section \ref{sec_CKT}, we then provide an introduction and generalization of CKT and the Wigner function formalism and present the quantum corrections on the energy-momentum transfer. In Section \ref{sec_chiral_transfer}, we derive the chiral radiation transport equation and Wigner functions of neutrinos in the inertial frame. Section \ref{sec_summary} is devoted to summary and outlook. 

Throughout this work, we assume massless neutrinos. We use the Minkowski metric $\eta_{\mu\nu}=\text{diag}\{+,-,-,-\}$. We define the Levi-Civita tensor $\epsilon^{\mu\nu\alpha\beta}=\hat{\epsilon}^{\mu\nu\alpha\beta}/\sqrt{-g}$, where $\hat{\epsilon}^{\mu\nu\alpha\beta}$ denotes the permutation symbol and $g$ represents the determinant of the spacetime metric with the convention $\hat{\epsilon}^{0123}=-\hat{\epsilon}_{0123}=1$. We absorb the electric charge $e$ into the definition of the gauge field $A_{\mu}$. We also introduce the notations $A_{\{\rho}B_{\sigma\}}\equiv (A_{\rho}B_{\sigma}+A_{\sigma}B_{\rho})/2$ and  $A_{[\rho}B_{\sigma]}\equiv (A_{\rho}B_{\sigma}-A_{\sigma}B_{\rho})/2$. We will keep $\hbar$ only to indicate the $\hbar$ expansion, but we will suppress other $\hbar$'s except for our main results in Equations~(\ref{Gamma_decomp})--(\ref{Gamma_omega}). We will take $c = 1$ after Section~\ref{sec_collision} and in Appendices~\ref{app_four-Fermi_theory}-\ref{app_conserv_eq}, except for Equations~(\ref{Gamma_decomp})--(\ref{Gamma_omega}).

\section{Classical radiation transport equation}
\label{sec_Boltzmann}
In this section, we review the derivation of the 3D classical transfer equation for delineating the neutrino radiation transport in the inertial frame. To make our discussion generic, we will first write down the Lorentz-covariant kinetic equation for charged particles in the presence of background electromagnetic fields in curvilinear coordinates. The kinetic equation for charge neutral neutrinos can be obtained by turning off the electromagnetic fields later.

We start with the Einstein-Vlasov equation, which is a generalized Boltzmann equation in curved spacetime or in non-Cartesian coordinates (non-Minkowski spacetime). For massless fermions, the Einstein-Vlasov equation reads
\begin{eqnarray}
	\delta(q^2)q\cdot \Delta f=0,
\end{eqnarray}
where $f(x, q)$ is the distribution function for a quasi-particle in phase space and
\begin{eqnarray}
	\Delta_{a}=\partial_{a}+(F^{c}_{\,\,\,a}-q^{b}\Gamma^{c}_{ab})\partial_{qc}
\end{eqnarray}
with $\partial_a\equiv \partial/\partial x^{a}$, $\Gamma^{c}_{ab}$ represents the Christoffel symbol, and $F_{ab}$ denotes the field strength for a U(1) gauge field. For the moment, we ignore the collisions on the right-hand side of the kinetic equation, which can be further included later. Note that here $q^a$ and $x^a$ in $f$ are independent, which is generally held in the off-shell case. Nevertheless, when implementing the on-shell condition, the derivatives with $x^{a}$ and with $q^{a}$ become entangled. To avoid the complexity, an efficient way is to introduce an orthonormal frame of local coordinates such that the $x^a$ and $q^a$ are independent under the on-shell condition. One then performs the corresponding coordinate transformation to the coordinate system $(x^{\mu},q^{\mu})$, e.g., $q^{a}=e^{a}_{\, \alpha}(x^{\mu})q^{\alpha}(x^{\mu})$ and $\eta_{ab}=e_{a}^{\,\alpha}(x^{\mu})e_{b}^{\,\beta}(x^{\mu})g_{\alpha \beta}(x^{\mu})$ via vierbeins \citep{Lindquist1966}. Here the Roman and Greek indices run over $\{0,1,2,3\}$ and $\{t,r,\theta,\phi\}$, respectively. 
Accordingly, for the Einstein-Vlasov equation, we have to apply the coordinate transformation on the Christoffel symbols,
\begin{eqnarray}
	\Gamma^{c}_{ab}=e^{c}_{\,\gamma}e_{a}^{\,\alpha} (e_{b}^{\,\beta}\Gamma^{\gamma}_{\alpha \beta}+\partial_{\alpha}e_b^{\,\gamma}),
\end{eqnarray}
which are called the Ricci rotation coefficients. Note that $q^a\Delta_a$ is invariant under the coordinate transformation while the individual terms $q^a\partial_a$ and $q^aq^b\Gamma^c_{ab}\partial_{qc}$ are not. 

We will now apply the equation above to obtain the renowned kinetic equation with spherically symmetric metric shown in \cite{Lindquist1966}. For generality, we will lift the spherical symmetry for the distribution functions and consider the general expression of a spherical symmetric spacetime metric, 
\begin{eqnarray}\label{s_metric}
	{\rm d}s^2={\rm e}^{2\Phi(t,\sr)}{\rm d}t^2-{\rm e}^{2\Lambda(t,\sr)}{\rm d}\sr^2-R(t,\sr)^2({\rm d}\theta^2+\sin^2\theta {\rm d}\phi^2),
\end{eqnarray}
which yields the following nonvanishing vierbeins, $e^0_{\, t}={\rm e}^{\Phi}$, $e^1_{\, \sr}={\rm e}^{\Lambda}$, $e^3_{\, \theta}=R$, and $e^4_{\, \phi}=R\sin\theta$. We also keep $\Phi$ and $\Lambda$ as arbitrary functions depending on $(t,{\sr})$ for generality.  
The corresponding four-momentum satisfying the null on-shell condition can be written as
\begin{eqnarray}\label{momentum_para}
	q^t={\rm e}^{-\Phi}E,\quad q^{\sr}={\rm e}^{-\Lambda}\mu E,\quad q^{\theta}=\frac{\sqrt{1-\mu^2}}{R}E\cos\bar{\phi}\,, \quad 
	q^{\phi}=\frac{\sqrt{1-\mu^2}}{R\sin\theta}E\sin\bar{\phi}\,,
\end{eqnarray}
where $\mu \equiv \cos \bar \theta$.
Note that here we only need three extra variables $(E,\mu,\bar{\phi})$ to parameterize $q^{\alpha}$ owing to the on-shell condition. 
Considering a general case for the distribution functions $f=f(t,\sr,\theta, \phi, E,\mu,\bar{\phi})$, the on-shell kinetic equation for charge neutral particles (when $F_{\mu\nu}=0$) reads
\begin{eqnarray}\nonumber
\label{classical_KT}
	0&=&\Big(q^{\alpha}\partial_{\alpha}-q^aq^b \Gamma^{c}_{ba} e_{c}^{\rho}\partial_{q\rho}\Big) f
	\\\nonumber
	&=&E\bigg(\tilde{\partial}_t+ \mu \tilde{\partial}_{\sr}+\frac{\sqrt{1-\mu^2}}{R}\cos\bar{\phi} \partial_{\theta}+\frac{\sqrt{1-\mu^2}}{R\sin\theta}\sin\bar{\phi}\partial_{\phi}
	-E\big(\mu \tilde{\partial}_{\sr} \Phi+(1-\mu^2)\tilde{\partial}_t\ln R+\mu^2\tilde{\partial}_t\Lambda\big)
	\partial_E 
	\\
	&&
	-(1-\mu^2)\big((\tilde{\partial}_{\sr}\Phi+\mu \tilde{\partial}_t\Lambda)-(\tilde{\partial}_{\sr}+\mu \tilde{\partial}_t)\ln R\big)\partial_{\mu}
	-\frac{\sqrt{1-\mu^2}}{R}\cot\theta\sin\bar{\phi} \partial_{\bar{\phi}}
	\bigg)f\,,
\end{eqnarray}
where $\tilde{\partial}_t \equiv {\rm e}^{-\Phi}\partial_t$ and $\tilde{\partial}_{\sr} \equiv {\rm e}^{-\Lambda}\partial_{\sr}$ and we used the relations in Equation~(\ref{der_rel}) shown in Appendix \ref{app_relations}. When further imposing the spherical symmetry for the distribution functions $f(t,r,E,\mu)$, the kinetic equation reduces to the one found in \citet{Lindquist1966}.  

We can directly implement Equation~(\ref{classical_KT}) to derive the kinetic equation in the inertial frame with the spacetime metric
\begin{eqnarray}\label{i_metric}
{\rm d}s^2=c^2{\rm d}t_{\rm i}^2-{\rm d}r^2-r^2({\rm d}\theta^2+\sin^2\theta{\rm d}\phi^2),
\end{eqnarray}
where the subscript ``i" represents the inertial frame and $c$ denotes the speed of light.
By comparing Equations~(\ref{s_metric}) and (\ref{i_metric}), we take
\begin{eqnarray}
	t= t_{\rm i},\quad \sr= r, \quad \Phi=\ln c, \quad\Lambda = 0,\quad R = r,
\end{eqnarray}
and we define the corresponding on-shell momentum,
\begin{eqnarray}\label{momentum_para_i}
	q^{t_{\rm i}}=\frac{E_{\rm i}}{c},\quad q^{r}=\mu_{\rm i} E_{\rm i},\quad q^{\theta}=\frac{\sqrt{1-\mu_{\rm i}^2}}{r}E_{\rm i}\cos\bar{\phi}_{\rm i} ,\quad 
	q^{\phi}=\frac{\sqrt{1-\mu_{\rm i}^2}}{r\sin\theta}E_{\rm i}\sin\bar{\phi}_{\rm i}\,.
\end{eqnarray}
Given $f=f(t_{\rm i},r,\theta, \phi, E_{\rm i},\mu_{\rm i},\bar{\phi}_{\rm i})$ in terms of the coordinates in the inertial frame, Equation (\ref{classical_KT}) reduces to
\begin{eqnarray}
\label{classical_KT_in}
	&&\left(\frac{1}{c}\partial_{t_{\rm i}}+ \mu_{\rm i}\partial_{r}+\frac{\sqrt{1-\mu_{\rm i}^2}}{r}\cos\bar{\phi}_{\rm i}\partial_{\theta}+ \frac{\sqrt{1-\mu_{\rm i}^2}}{r\sin\theta}\sin\bar{\phi}_{\rm i}\partial_{\phi}
	+\frac{1-\mu_{\rm i}^2}{r}\partial_{\mu_{\rm i}} - \frac{\sqrt{1-\mu_{\rm i}^2}}{r}\sin\bar{\phi}_{\rm i} \cot\theta \partial_{\bar{\phi}_{\rm i}}\right)f = 0\,,
\end{eqnarray}
which does not depend on the fluid velocity and the energy derivative. For numerical calculations, it is practical to rewrite the transfer equations into a conservative form. The conservative form of Equation~(\ref{classical_KT_in}) becomes
\begin{eqnarray}
\label{classical_KT_in_con}
\Bigg[\frac{1}{c}\partial_{t_{\rm i}}+ \frac{\mu_{\rm i}}{r^2}\partial_{r}r^2+\frac{\sqrt{1-\mu_{\rm i}^2}}{r}\Big(\frac{\cos\bar{\phi}_{\rm i}}{\sin\theta}\partial_{\theta}\sin\theta+\frac{\sin\bar{\phi}_{\rm i}}{\sin\theta}\partial_{\phi}\Big)
 +\frac{1}{r}\partial_{\mu_{\rm i}}(1-\mu_{\rm i}^2)
 -\frac{\sqrt{1-\mu_{\rm i}^2}}{r}\cot\theta\partial_{\bar{\phi}_{\rm i}}\sin\bar{\phi}_{\rm i}
 \Bigg]f = 0 \,.
 \end{eqnarray}
This expression can also be found in, e.g., \citet{Sumiyoshi:2012za}. Finally, one has to retrieve the collision terms responsible for radiation transfer in Equations~(\ref{classical_KT_in_con}), which will be discussed later with the inclusion of quantum corrections.

\section{Chiral kinetic theory}
\label{sec_CKT}
\subsection{Wigner Functions and Kinetic Theory}\label{WF_CKT}
In this section, we shortly review and generalize the CKT for massless chiral fermions obtained from the Wigner function approach in curved spacetime. For generality, we will first consider charged particles in the presence of electromagnetic fields again, but we will focus on charge neutral neutrinos by turning off the electromagnetic fields later. As a starting point, we introduce the Wigner functions for left-handed fermions as the quantum expectation values of correlation functions in Minkowski spacetime,%
\footnote{The exponential factor ${\rm e}^{-\frac{{\rm i}q\cdot y}{\hbar c}}$ in Equation (\ref{def_WF_f}) may look different from the one, ${\rm e}^{-\frac{{\rm i}q\cdot y}{\hbar}}$, in the usual field theory literature. This originates from the fact that, in this paper, we follow the convention of coordinates, $q^{\alpha} = (t, \sr, \theta, \phi)$, in the literature of radiation hydrodynamics \citep[e.g.,][]{Mihalas} unlike the convention $q^{\alpha} = (ct, \sr, \theta, \phi)$ of the field theory literature.}
\begin{eqnarray}
\label{def_WF_f}
	\grave{S}_{\rm L}^{\lessgtr}(q,x)\equiv\int {\rm d}^4y \ {\rm e}^{-\frac{{\rm i}q\cdot y}{\hbar c}}S_{\rm L}^{\lessgtr}(x,y)\,,
\end{eqnarray}
where $S_{\rm L}^<(x,y) \equiv \langle \psi^{\dagger}_{\rm L}(x+y/2)\psi_{\rm L}(x-y/2)\rangle$ and $S_{\rm L}^>(x,y) \equiv \langle\psi_{\rm L}(x-y/2){\psi}^{\dagger}_{\rm L}(x+y/2)\rangle$ are the lesser and greater propagators for left-handed fermions, respectively \citep[see, e.g.,][for a review]{Blaizot:2001nr}. Here left- and right-handed fermions $\psi_{\rm L,R}$ are defined as $\psi_{\rm L,R}\equiv P_{\rm L,R} \psi$ for a Dirac fermion $\psi$, with the projection operators $P_{\rm L,R} \equiv (1 \mp \gamma^5)/2$ and $\gamma^5={\rm i}\gamma^0\gamma^1\gamma^2\gamma^3$. Between the field operators $\psi^{\dagger}_{\rm L}$ and $\psi_{\rm L}$ in the expressions above, gauge links are implicitly embedded to preserve gauge invariance. The dynamics of Wigner functions in phase space are then dictated by Kadanoff-Baym equations derived from the Dirac equation. Nevertheless, in order to solve Kadanoff-Baym equations, one has to further perform the $\hbar$ expansion that is equivalent to a gradient expansion. One then perturbatively solves the Kadanoff-Baym equations for Wigner functions with the $\hbar$ expansion up to $O(\hbar)$ to capture the leading-order quantum corrections and thereby derives the corresponding CKT as a modified Boltzmann equation \citep{Hidaka:2016yjf}.      

In curved spacetime, the definition of phase space becomes more subtle owing to the lack of global momentum. Instead, the phase space is defined on a tangent or cotangent bundle as applied in \citet{Liu:2018xip} for the derivation of CKT in curve spacetime (or more precisely, non-Minkowski spacetime there). For convenience, we will choose the tangent bundle with the set $(x^{\mu},q^{\mu})$ as opposed to the choice in \citet{Liu:2018xip}. The Wigner functions and CKT may differ, but the physics remain unchanged when making different choices. Now, the definition of Wigner functions becomes
\begin{eqnarray}\label{def_WF_c}
\grave{S}_{\rm L}^{\lessgtr}(q,x)\equiv\int \frac{{\rm d}^4y}{\sqrt{-g(x)}} {\rm e}^{-\frac{{\rm i}q\cdot y}{\hbar c}}S_{\rm L}^{\lessgtr}(x,y)\,,
\end{eqnarray}
where $S_{\rm L}^<(x,y) \equiv \langle \psi^{\dagger}_{\rm L}(x, y/2)\psi_{\rm L}(x, -y/2)\rangle$ and $S_{\rm L}^>(x,y) \equiv \langle\psi_{\rm L}(x, -y/2)\psi^{\dagger}_{\rm L}(x, y/2)\rangle$ and $g(x)$ denotes the determinant of the spacetime metric. Here $\psi_{\rm L}(x,y)={\rm e}^{y\cdot\tilde{D}}\psi_{\rm L}(x)$ and $\psi^{\dagger}_{\rm L}(x,y)=\psi^{\dagger}_{\rm L}(x){\rm e}^{y\cdot\overleftarrow{\tilde{D}}}$, where $\tilde{D}_{\mu}=\nabla_{\mu}+{\rm i}A_{\mu}/\hbar+\Gamma^{\lambda}_{\mu\nu}y_{\lambda}\partial_y^{\nu}$ corresponds to the horizontal lift and $\nabla_{\mu}$ denotes the covariant derivative with respect to $x^{\mu}$. It turns out that the horizontal lift provides a proper covariant derivative on the phase space such that $\tilde{D}_{\mu}y_{\nu}=0$ and $\tilde{D}_{\mu}g_{\alpha\beta}(x)=0$ when $A_{\mu}=0$. With this definition, Equation~(\ref{def_WF_c}) reduces to Equation~(\ref{def_WF_f}) in Minkowski spacetime. Despite the technical subtleties, the strategy for the derivation of CKT in the Wigner function formalism in curved spacetime is the same as that in Minkowski spacetime. One may refer to \citet{Liu:2018xip} for more details. The lesser propagator of left-handed fermions can be parameterized as $\grave{S}_{\rm L}^{\lessgtr}(q,x)=\sigma^{\mu}\mathcal{L}^{\lessgtr}_{\mu}(q,x)$, where $\sigma^{\mu} = ({\bm I}, \sigma^1, \sigma^2, \sigma^3)$ with ${\bm I}$ being an identity matrix and $\sigma^1$, $\sigma^2$, $\sigma^3$ the Pauli matrices.%
\footnote{One can in fact construct the Wigner functions for Dirac fermions, $S^{<}(q,x)$, by replacing $S^{<}_{\rm L}(x,y)$ in Equation~(\ref{def_WF_c}) with $S^<(x,y)=\langle \bar{\psi}(x,y/2)\psi(x,-y/2)\rangle$. Based on the Clifford algebra, one may decompose the Wigner functions as, e.g., $\grave{S}^<=\mathcal{S}+ {\rm i}\mathcal{P}\gamma^5+ \mathcal{V}^{\mu}\gamma_\mu+\mathcal{A}^{\mu}\gamma^5\gamma_{\mu}+ \frac{\mathcal{S}^{\mu\nu}}{2}\Sigma_{\mu\nu}$, where $\Sigma_{\mu\nu}={\rm i}[\gamma_{\mu},\gamma_{\nu}]/2$ \citep{Vasak:1987um}. In the massless limit, $\mathcal{V}^{\mu}$ and $\mathcal{A}^{\mu}$ are decoupled from $\mathcal{S}$, $\mathcal{P}$, and $\mathcal{S}^{\mu\nu}$. One may further define $\grave{S}^<_{\rm R}=P_{\rm R}\gamma^{\mu}\mathcal{R}_{\mu}$ and $\grave{S}^<_{\rm L}=P_{\rm L}\gamma^{\mu}\mathcal{L}_{\mu}$ with $\grave{S}_{\rm R/L}^<$ being the lesser propagators of right-handed/left-handed fermions. In the Weyl basis, one finds $\grave{S}^<_{\rm L}=\sigma^{\mu}\mathcal{L}^<_{\mu}$.} 

However, in order to construct the radiation hydrodynamic incorporating the energy-momentum transfer between neutrinos and matter, it is inevitable to include collision terms, which are not considered in \citet{Liu:2018xip}. Although a rigorous derivation of collisions in the CKT in curved spacetime might be technically more involved, we may generalize the derivation of the CKT with collisions in Minkowski spacetime shown in \cite{Hidaka:2016yjf,Hidaka:2017auj} with proper modifications upon the Kadanoff-Baym equation to the case of curve spacetime. In light of the approach in \citet{Hidaka:2016yjf,Hidaka:2017auj,Liu:2018xip}, the Kadanoff-Baym equation with collisions for left-handed fermions leads to the following master equations up to $O(\hbar)$:%
\footnote{ Here and below, we ignore the one-particle potential denoted by $\Sigma^{\delta}$ in \citet{Hidaka:2016yjf} for simplicity, as it is irrelevant to the chiral effects that we are interested in. The inclusion of $\Sigma^{\delta}$ may modify the dispersion relation of the fermions.}
\begin{eqnarray}\label{m_eq1}
	\mathcal{D}\cdot \mathcal{L}^{<}&=&0,
	\\\label{m_eq2}
	q\cdot\mathcal{L}^{<}&=&0,
	\\\label{m_eq3}
	\hbar c \big(\mathcal{D}_{\mu}\mathcal{L}^{<}_{\nu}-\mathcal{D}_{\nu}\mathcal{L}^{<}_{\mu}\big)&=&-2\epsilon_{\mu\nu\rho\sigma}q^{\rho}\mathcal{L}^{< \sigma},
\end{eqnarray}
where 
\begin{eqnarray}
\mathcal{D}_{\mu}\mathcal{L}_{\nu}^< \equiv \Delta_{\mu}\mathcal{L}^<_{\nu}-\Sigma^<_{\mu}\mathcal{L}^>_{\nu}+\Sigma^>_{\mu}\mathcal{L}^<_{\nu}
\end{eqnarray}
and $\Delta_{\mu}\mathcal{L}_{\nu} \equiv \big(D_{\mu}+F_{\lambda\mu}\partial^{\lambda}_q\big)\mathcal{L}_{\nu}$ with $D_{\mu} \equiv \nabla_{\mu}-\Gamma^{\lambda}_{\mu\nu}q^{\nu}\partial_{q\lambda}$. Recall that $\nabla_{\mu}$ denotes the covariant derivative with respect to $x^{\mu}$ such that $\nabla_{\mu}\mathcal{L}_{\nu}=\partial_{\mu}\mathcal{L}_{\nu}-\Gamma^{\lambda}_{\mu\nu}\mathcal{L}_{\lambda}$. Note that $D_{\mu}q^{\nu}=0$, where $D_{\mu}$ is the dual operator of the horizontal lift $\tilde{D}_{\mu}$ in the tangent space $(x^{\mu},q^{\mu})$ when neglecting gauge fields. Also, $\Sigma^{<}_{\mu}$ and $\Sigma^{>}_{\mu}$ correspond to lesser and greater self-energies depending on details of interactions in a given system. Here $\hbar$ can be regarded as an expansion parameter to track the quantum corrections. 
Equation~(\ref{m_eq1}) is constructed by replacing the spacetime derivatives $\partial_{\mu}$ by $D_{\mu}$ in the master equations in \cite{Hidaka:2016yjf,Hidaka:2017auj}.
From Equations~(\ref{m_eq2}) and (\ref{m_eq3}), the corresponding solution up to $O(\hbar)$ takes the form (see Appendix \ref{app_Wigner})%
\footnote{We here ignored the contribution of antiparticles, which can be included by multiplying the right-hand side of Equation~(\ref{WF_L_full}) by the sign of $q\cdot n$.}
\begin{eqnarray}\label{WF_L_full}
	\mathcal{L}^{\lessgtr \mu}=2\pi\Big[\delta(q^2)\big(q^{\mu}-\hbar c S^{\mu\nu}_{(n)}\mathcal{D}_{\nu}\big)-\hbar c \tilde{F}^{\mu\nu}q_{\nu}\delta'(q^2)\Big]f_{\rm L}^{\lessgtr}\,,
\end{eqnarray} 
where $\mathcal{D}_{\mu}f_{\rm L} \equiv D_{\mu}f_{\rm L}-\mathcal{C}_{\mu}[f_{\rm L}]$ and $\mathcal{C}_{\mu} [f_{\rm L}^{\lessgtr}] \equiv\Sigma_{\mu}^{\lessgtr}f^{\gtrless}_{\rm L}-\Sigma_{\mu}^{\gtrless}f^{\lessgtr}_{\rm L}$, with $f^{<}_{\rm L}=f_{\rm L}$ and $f^{>}_{\rm L}=1-f_{\rm L}$ the distribution functions of incoming and outgoing fermions, respectively. Here
\begin{eqnarray}
	S^{\mu\nu}_{(n)}=\frac{\epsilon^{\mu\nu\alpha\beta}q_{\alpha}n_{\beta}}{2q\cdot n}
\end{eqnarray}
denotes the spin tensor, which depends on a timelike frame vector $n^{\mu}(x)$ satisfying $n^2=1$ and $\delta'(q^2)=\partial\delta(q^2)/\partial q^2$ and $\tilde{F}^{\mu\nu}=\epsilon^{\mu\nu\alpha\beta}F_{\alpha\beta}/2$. The frame vector $n^{\mu}(x)$ appears as a choice of the spin basis such that $n^{\mu}\sigma_{\mu}={\bm I}$ and that $\sigma^{\mu}_{\perp}$ perpendicular to $n^{\mu}$ becomes $\sigma^{\mu}_{\perp} = (0, \sigma^1, \sigma^2, \sigma^3)$. That is, we define $n^{\mu}=e^{\mu}_0$ as the zeroth component of vierbeins \citep{Hidaka:2018ekt}. However, one should note that $\mathcal{L}^{\mu}$ is independent of the choice of $n^{\mu}$. In addition, as discussed in \citet{Hidaka:2016yjf}, the $O(\hbar)$ corrections proportional to $q^{\mu}\delta(q^2)$ as the trivial solutions for Equations~(\ref{m_eq2}) and (\ref{m_eq3}) can be absorbed into $f_{\rm L}$.

The quantum corrections at $O(\hbar)$ now incorporate two terms shown in Equation~(\ref{WF_L_full}), in which the $\delta'(q^2)$ term yields the modification on the on-shell condition due to the magnetic-moment coupling in the presence of background electromagnetic fields \citep{Son:2012zy,Chen:2014cla}. Although such a term vanishes for neutrinos, the other term in Equation~(\ref{WF_L_full}) associated with the spin tensor $S^{\mu\nu}_{(n)}$, called the side-jump term, exists even without background electromagnetic fields, which then influences the neutrino transport. This side-jump term stems from the spin-momentum locking of chiral fermions under the angular-momentum conservation, and it contributes to the so-called magnetization currents and CVE \citep{Chen:2014cla,Chen:2015gta,Yang:2018lew}. 

Note that the $n^{\mu}$ dependence of $S^{\mu\nu}_{(n)}$ implies that $f_{\rm L}$ is no longer invariant under the frame transformation. Given the fact that $\mathcal{L}^{\mu}$ is frame independent, one can accordingly derive the modified frame transformation on $f_{\rm L}$ between different frame choices, which is also related to the modified Lorentz transformation  \citep{Chen:2014cla,Chen:2015gta,Hidaka:2016yjf}. More precisely, the distribution function $f_{\rm L}^{(n)}$ in one frame with $n^{\mu}$ is related to $f_{\rm L}^{(n')}$ in another frame with $n'^{\mu}$ by
\begin{eqnarray}
\label{frame_transf}
f_{\rm L}^{(n')}= f_{\rm L}^{(n)}-\hbar c \frac{\epsilon^{\nu\mu\alpha\beta}q_{\alpha}n'_{\beta}n_{\mu}}{2(q\cdot n)(q\cdot n')}\mathcal{D}_{\nu}f_{\rm L}^{(n)}\,.
\end{eqnarray}
Note that the frame transformation between different frames is distinct from the usual coordinate transformation between the inertial frame and the comoving frame in radiation hydrodynamics. The two different notions of these ``frame transformations" should not be confused with each other.  

By plugging Equation~(\ref{WF_L_full}) into Equation~(\ref{m_eq1}) and employing the relation
\begin{eqnarray}\nonumber
	[\Delta_{\mu},\Delta_{\nu}]f_{\rm L}&=&\big[\big(\nabla_{\mu}+F_{\lambda\mu}\partial^{\lambda}_q-\Gamma^{\lambda}_{\mu\rho}q^{\rho}\partial_{q\lambda}\big),\big(\nabla_{\nu}+F_{\lambda'\nu}\partial^{\lambda'}_q-\Gamma^{\lambda'}_{\nu\rho'}q^{\rho'}\partial_{q\lambda'}\big)\big]f_{\rm L}
	\\\nonumber
	&=&\Big[2(\nabla_{[\mu}F_{\lambda\nu]})\partial_q^{\lambda}-2q^{\rho}\big(\nabla_{[\mu}\Gamma^{\lambda}_{\nu]\rho}\partial_{q\lambda}\big)\Big]f_{\rm L}
	\\
	&=&\Big[2(\nabla_{[\mu}F_{\lambda\nu]})\partial_q^{\lambda}-q^{\rho}R^{\lambda}_{\rho\mu\nu}\partial_{q\lambda}\Big]f_{\rm L}
	\,,
\end{eqnarray}
where we used $R^{\lambda}_{\rho\mu\nu}=2\partial_{[\mu}\Gamma^{\lambda}_{\nu]\rho}+2\Gamma^{\lambda}_{\alpha[\mu}\Gamma^{\alpha}_{\nu]\rho}$, the CKT in curved spacetime as a modified Einstein-Vlasov equation up to $O(\hbar)$ is derived as
\begin{eqnarray}
	\delta\big(q^2-\hbar c F_{\alpha\beta}S^{\alpha\beta}_{(n)}\big)\Bigg[q\cdot\tilde{\mathcal{D}}-\hbar c \left(\frac{S^{\mu\nu}_{(n)}F_{\mu\rho}n^{\rho}}{q\cdot n}+\big(D_{\mu}S^{\mu\nu}_{(n)}\big)\right) \mathcal{D}_{\nu}
	-\hbar c S^{\mu\nu}_{(n)}\Big(\nabla_{\mu}F^{\lambda}_{\,\,\,\nu}-q^{\rho}R^{\lambda}_{\rho\mu\nu}\Big)\partial_{q\lambda}
	\Bigg]f_{\rm L}=0\,,
\end{eqnarray}
where $\tilde{\mathcal{D}}_{\mu}f_{\rm L} \equiv \Delta_{\mu} f_{\rm L}-\tilde{\mathcal{C}}_{\mu}[f_{\rm L}]$ and
\begin{eqnarray}
\tilde{\mathcal{C}}^{\mu}[f_{\rm L}] \equiv \mathcal{C}^{\mu}[f_{\rm L}]-\hbar c\frac{\epsilon^{\mu\nu\alpha\beta}n_{\nu}}{2q\cdot n}
\big((1-f_{\rm L})\Delta^>_{\alpha}\Sigma^{<}_{\beta}-f_{\rm L}\Delta^<_{\alpha}\Sigma^{>}_{\beta}\big)
\end{eqnarray} 
with $\Delta^{\gtrless}_{\mu} \equiv \Delta_{\mu}+\Sigma_{\mu}^{\gtrless}$. For right-handed fermions, the $O(\hbar)$ terms flip signs. 

To delineate the transport for neutrinos, we can turn off the background electromagnetic fields. The dispersion relation for chiral fermions hence remains lightlike. In the flat spacetime such as in the inertial frame, we can further drop the term proportional to the Riemann tensor. It turns out that only the term associated with the horizontal lift acting on the spin tensor contributes to the quantum corrections. We can explicitly evaluate this term,
\begin{eqnarray}
	D_{\mu}S^{\mu\nu}_{(n)}=
	\frac{\epsilon^{\mu\nu\alpha\beta}q_{\alpha}}{2q\cdot n}\left(\nabla_{\mu}n_{\beta}-n_{\beta}q^{\rho}\frac{\nabla_{\mu}n_{\rho}}{q\cdot n}\right)\,,
\end{eqnarray}  
where we applied the property of the Levi-Civita tensor, $\nabla_{\rho}\epsilon^{\mu\nu\alpha\beta}=0$, and $D_{\mu}q^{\nu}=0$. For generality, we will assume a nonvanishing Riemann tensor, in which case the CKT for left-handed neutrinos is given by
\begin{eqnarray}
	&&\left[q^{\mu}(\partial_{\mu}-\Gamma^{\lambda}_{\mu\rho}q^{\rho}\partial_{q\lambda})
	-\hbar c \frac{\epsilon^{\mu\nu\alpha\beta}q_{\alpha}}{2q\cdot n}\left(\nabla_{\mu}n_{\beta}-n_{\beta}q^{\rho}\frac{\nabla_{\mu}n_{\rho}}{q\cdot n}\right)\partial_{\nu}
	+\hbar c \frac{\epsilon^{\mu\nu\alpha\beta}q_{\alpha}n_{\beta}}{2q\cdot n}q^{\rho}R^{\lambda}_{\rho\mu\nu}\partial_{q\lambda}\right]f_{\rm L}
	\nonumber \\ 
	\label{CKT_noBF}
	&& 
	=(1-f_{\rm L}) \Gamma^{<}_{(n)} - f_{\rm L} \Gamma^{>}_{(n)}\,,
\end{eqnarray}
under the on-shell condition $q^2 =0$, where
\begin{eqnarray}
\label{Gamma}
\Gamma^{\lessgtr}_{(n)} = q \cdot \Sigma^{\lessgtr} - \hbar c \frac{\epsilon^{\mu\nu\alpha\beta}q_{\alpha}}{2q\cdot n} 
\left[ \left(\nabla_{\mu}n_{\beta}-n_{\beta}q^{\rho}\frac{\nabla_{\mu}n_{\rho}}{q\cdot n}\right) \Sigma^{\lessgtr}_{\nu} + n_{\beta} D_{\mu}\Sigma^{\lessgtr}_{\nu} \right]\,
\end{eqnarray}
are related to the emission and absorption rates via $R_{\rm emis} = c\Gamma^{<}/E$ and $R_{\rm abs} = c\Gamma^{>}/E$, respectively.
Here we also dropped the nonlinear terms in self-energies due to the weakness of the weak interaction.

Note that all the quantum corrections in Equations (\ref{CKT_noBF}) and (\ref{Gamma}) involve the Levi-Civita tensor $\epsilon^{\mu\nu\alpha\beta}$, and hence they explicitly break the spherical symmetry and axisymmetry of the system. This consequence may simply be understood from the fact that the chirality related to the spin degree of freedom can only be defined in genuine 3D.

\subsection{Conservative Equations and Energy-Momentum Transfer}
After solving the CKT and obtaining $f_{\rm L}$, we have to insert $f_{\rm L}$ into the Wigner function in Equation~(\ref{WF_L_full}) to obtain physical observables. For example, based on the definition in field theory, the particle-number current and symmetric energy-momentum tensor for left-handed fermions can be derived from the lesser propagators via 
\begin{eqnarray}\label{J_and_T}
J^{\mu}=2\int_q\mathcal{L}^{<\mu},
\qquad
T^{\mu\nu}=\int_q \big(\mathcal{L}^{<\mu}q^{\nu}+\mathcal{L}^{<\nu}q^{\mu}\big),
\end{eqnarray}
where we introduced the notation
\begin{eqnarray}
\int_q \equiv \int\frac{{\rm d}^4q}{(2\pi)^4}\sqrt{-g}\,.
\end{eqnarray}
From Equations~(\ref{J_and_T}) and (\ref{WF_L_full}), in the absence of $\hbar$ corrections, one easily recognizes that $J^{\mu}$ and $T^{\mu\nu}$ simply reduce to taking the first and second moments of $f_{\rm L}$, respectively. 

In fact, the quantum corrections further affect the conservative equations responsible for the energy-momentum transfer between neutrinos and matter. As shown in Appendix~\ref{app_conservative}, the conservative equation for the radiation energy-momentum tensor of neutrinos is given by
\begin{eqnarray}\label{cons_Tmunu}
\nabla_{\mu}T^{\mu\nu}_{\text{rad}}
=2\int_q \left(q^{\nu}\check{\mathcal{C}}[f_{\rm L}]
+\frac{\pi}{2}\hbar c\delta(q^2)\epsilon^{\nu\mu\alpha\beta}q_{\alpha}D_{\beta}\mathcal{C}_{\mu}[f_{\rm L}]\right),
\end{eqnarray}
where 
\begin{eqnarray}
\label{C_check}
\check{\mathcal{C}}[f_{\rm L}] \equiv \Sigma^<\cdot\mathcal{L}^>-\Sigma^>\cdot\mathcal{L}^<=
2\pi\delta(q^2)\left[ q\cdot\mathcal{C}-\hbar c\frac{\epsilon^{\mu\rho\alpha\beta}q_{\alpha}n_{\beta}}{2q\cdot n}\Big(\Sigma^<_{\mu}D_{\rho}(1-f_{\rm L})-\Sigma^>_{\mu}D_{\rho}f_{\rm L}\Big)\right]\,.
\end{eqnarray}
Based on the energy-momentum conservation, 
\begin{eqnarray}
\nabla_{\mu}T^{\mu\nu}_{\text{rad}}+\nabla_{\mu}T^{\mu\nu}_{\text{mat}}=0,
\end{eqnarray}  
where $T^{\mu\nu}_{\text{mat}}$ denotes the energy-momentum tensor of matter, the right-handed side of Equation~(\ref{cons_Tmunu}) will accordingly modify the transport of matter. Note that $\Sigma^{\lessgtr}_{\mu}$ can also incorporate quantum corrections.

\section{Chiral radiation transport equation for neutrinos}
\label{sec_chiral_transfer}
\subsection{Transfer Equation in the Inertial Frame}
In this section, we will further write down an explicit expression of the transfer equation for left-handed neutrinos including quantum corrections in the inertial frame with the spacetime metric in Equation~(\ref{i_metric}). Thanks to the unchanging lightlike dispersion relation, we can still apply the same momentum parameterization as Equation~(\ref{momentum_para}). Nevertheless, we have to choose a proper ``frame vector $n^{\mu}$" for computational convenience, yet the physics should be independent of the choice. In the inertial frame, we may take $n^{\mu}=\xi^{\mu} \equiv (1/c,0,0,0)$ such that $q\cdot n=E_{\rm i}$. One may in general choose an arbitrary timelike frame vector for $n^{\mu}$, which, however, may cause unnecessary complication in calculations when solving for $f_{\rm L}^{(n)}$. Recall that one can always relate the distribution functions in different frames through the ``modified frame transformation" in Equation~(\ref{frame_transf}). Moreover, since the collision terms involve incoming and outgoing particles with different momenta, we will hereafter use extra subscripts to characterize the momentum dependence of variables and operators (e.g., $f^{(n)}_p\equiv f^{(n)}(p,x)$ and $D_{p\mu}\equiv \nabla_{\mu}-\Gamma^{\lambda}_{\mu\nu}p^{\nu}\partial_{p\lambda}$).  

Taking $n^{\mu}=\xi^{\mu}$ in the inertial frame, we have $\nabla_{\mu}n_{\nu}=0$, $D_{\mu}S^{\mu\nu}_{(n)}=0$, and $R^{\lambda}_{\rho\mu\nu}=0$, from which we find that the free-streaming part of the transfer equation remains unchanged from Equations~(\ref{classical_KT_in}) and (\ref{classical_KT_in_con}). Moreover, the collision terms are simplified. The transfer equation in the conservative form with collisions is thus given by
 \begin{eqnarray}
\label{CKT_inertial_con}
\nonumber
&&\Bigg[\frac{1}{c}\partial_{t_{\rm i}}+ \frac{\mu_{\rm i}}{r^2}\partial_{r}r^2+\frac{\sqrt{1-\mu_{\rm i}^2}}{r}\Big(\frac{\cos\bar{\phi}_{\rm i}}{\sin\theta}\partial_{\theta}\sin\theta+\frac{\sin\bar{\phi}_{\rm i}}{\sin\theta}\partial_{\phi}\Big)
 +\frac{1}{r}\partial_{\mu_{\rm i}}(1-\mu_{\rm i}^2)
 -\frac{\sqrt{1-\mu_{\rm i}^2}}{r}\cot\theta\partial_{\bar{\phi}_{\rm i}}\sin\bar{\phi}_{\rm i}
 \Bigg]f_{{\rm L}q}^{(\xi)}
 \\
 &&=\frac{1}{E_{\rm i}}\left[(1-f^{(\xi)}_{{\rm L}q})\Gamma^{<}_{(\xi)q}-f_{{\rm L}q}^{(\xi)}\Gamma^{>}_{(\xi)q} \right]\,,
 \end{eqnarray}
where, according to Equation~(\ref{Gamma}), 
\begin{eqnarray}
\label{Gamma_inertial}
\Gamma^{\lessgtr}_{(\xi)q} = \Big(q^{\nu}-\hbar cS^{\mu\nu}_{(\xi)q}D^{({\rm i})}_{q\mu}\Big)\Sigma^{\lessgtr}_{q\nu}\,,
\end{eqnarray}
and
\begin{eqnarray}
\nonumber
D^{({\rm i})}_{q\nu}&=&\nabla_{\nu}-\big(\Gamma^{\lambda}_{\nu\sigma}+e^a_{\,\,\sigma} \partial_{\nu}e^{\,\,\lambda}_a \big)q^{\sigma}\partial_{q\lambda}
\\\nonumber
&=&\Bigg(\nabla_{t_{\rm i}},\nabla_{r},\nabla_{\theta}+\frac{1}{\sqrt{1-\mu_{\rm i}^2}}\big((1-\mu_{\rm i}^2)\cos\bar{\phi}_{\rm i}\partial_{\mu_{\rm i}}+\mu_{\rm i}\sin\bar{\phi}_{\rm i}\partial_{\bar{\phi}_{\rm i}}\big),
\\
&&\quad
\nabla_{\phi}-\cos\theta\partial_{\bar{\phi}_{\rm i}}-\frac{\sin\theta}{\sqrt{1-\mu_{\rm i}^2}}
\big(\mu_{\rm i}\cos\bar{\phi}_{\rm i}\partial_{\bar{\phi}_{\rm i}}-(1-\mu_{\rm i}^2)\sin\bar{\phi}_{\rm i}\partial_{\mu_{\rm i}}\big)
\Bigg)\,.
\end{eqnarray}
Note again that the horizontal lifts here entail the proper coordinate transformation for Christoffel symbols to handle the mixing of spatial and momentum derivatives when working in the coordinates with on-shell momenta. Since $\Sigma_{\nu}^{\lessgtr}$ further contain quantum corrections, we have not written down the explicit expression for collisions above. The collision terms depend on microscopic theories or models characterizing the interactions between neutrinos and matter, which will be further discussed in the next subsection. For convenience, we will hereafter suppress the superscript ``$(\xi)$" for $f^{(\xi)}_{\rm L}$.

Moreover, the side-jump term also affects the neutrino radiation through the radiative energy-momentum tensor via the Wigner functions in Equation~(\ref{WF_L_full}). Unlike the classical case, in particular, the energy-momentum tensor no longer corresponds to the second moment of $f_{\rm L}$ in the presence of quantum corrections as mentioned in Section \ref{sec_CKT}. After solving $f_{\rm L}$ from the transfer equations, we have to further employ the Wigner functions to evaluate physical observables for neutrino radiation. In the case of neutrinos, we can turn off the background electromagnetic fields. Similarly to the transfer equation, the proper coordinate transformation for Christoffel symbols has to be taken care of for $D_{\nu}$. Following Section \ref{sec_Boltzmann}, we find
	\begin{eqnarray}
	\nonumber
		\mathcal{L}^{\mu}_q
		&=&2\pi\delta(q^2)e^{\,\,\mu}_d\Big[q^df_{{\rm L}q}-\hbar c S^{db}_{(n)q}\big(\partial_b-\Gamma^{c}_{ba}q^a\partial_{qc}\big)f_{{\rm L}q}+\hbar c S^{db}_{(n)q}\mathcal{C}_{qb}\Big]
		\\
		&=&2\pi\delta(q^2)\Big[q^{\mu}f_{{\rm L}q}-\hbar c S^{\mu\nu}_{(n)q}\Big(\partial_{\nu}f_{{\rm L}q}-\big(\Gamma^{\lambda}_{\nu\sigma}+e^a_{\,\,\sigma} \partial_{\nu}e^{\,\,\lambda}_a \big)q^{\sigma}\partial_{q\lambda}\Big)f_{{\rm L}q}
		+\hbar cS^{\mu\nu}_{(n)q}\mathcal{C}_{q\nu}
		\Big]\,.
	\end{eqnarray}
Defining $\mathcal{L}^{\mu}_q\equiv 2\pi\delta(q^2)\hat{\mathcal{L}}_q^{\mu}f_{{\rm L}q}$ and carrying out a straightforward computation, we obtain 
	\begin{eqnarray}\label{Lf_inetrial}
	\nonumber
	\hat{\mathcal{L}}^{t_{\rm i}}_qf_{{\rm L}q}&=&\frac{E_{\rm i}}{c}f_{{\rm L}q}\,,
	\\\nonumber
	\hat{\mathcal{L}}^{r}_qf_{{\rm L}q}&=&\mu_{\rm i} E_{\rm i}f_{{\rm L}q}
	-\frac{\hbar c}{2r} \bigg[ \sqrt{1-\mu_{\rm i}^2}\Big(\sin\bar{\phi}_{\rm i}\big(\partial_{\theta}f_{{\rm L}q}-\mathcal{C}_{q\theta}\big) - \frac{\cos\bar{\phi}_{\rm i}}{\sin\theta}\big(\partial_{\phi}f_{{\rm L}q}-\mathcal{C}_{q\phi}\big)\Big)
		+\Big(\sqrt{1-\mu_{\rm i}^2} \cot\theta \cos\bar{\phi}_{\rm i} +\mu_{\rm i}\Big)\partial_{\bar{\phi}_{\rm i}}f_{{\rm L}q}
		\bigg]\,,
	\\\nonumber
	\hat{\mathcal{L}}_q^{\theta}f_{{\rm L}q}&=&\frac{\sqrt{1-\mu_{\rm i}^2}}{r}E_{\rm i}\cos\bar{\phi}_{\rm i}f_{{\rm L}q}
+\frac{\hbar c}{2r^2}\bigg[\sqrt{1-\mu_{\rm i}^2} r \sin\bar{\phi}_{\rm i}\big(\partial_{r}f_{{\rm L}q}-\mathcal{C}_{qr}\big)-\frac{\mu_{\rm i}}{\sin \theta}\big(\partial_{\phi}f_{{\rm L}q}-\mathcal{C}_{q\phi}\big)-\mu_{\rm i}\sqrt{1-\mu_{\rm i}^2}\sin\bar{\phi}_{\rm i}\partial_{\mu_{\rm i}}f_{{\rm L}q}
\\\nonumber
&&+\mu_{\rm i}\bigg(\frac{\mu_{\rm i}\cos\bar{\phi}_{\rm i}}{\sqrt{1-\mu_{\rm i}^2}}+\cot\theta\bigg)\partial_{\bar{\phi}_{\rm i}}f_{{\rm L}q}\bigg]\,,
\\\nonumber
	\hat{\mathcal{L}}_q^{\phi}f_{{\rm L}q}&=&\frac{\sqrt{1-\mu_{\rm i}^2}}{r\sin\theta}E_{\rm i}\sin\bar{\phi}_{\rm i}f_{{\rm L}q}-
	\frac{\hbar c}{2 r^2 \sin \theta}
	\bigg[\sqrt{1-\mu_{\rm i}^2}\cos\bar{\phi}_{\rm i}r\big(\partial_{r}f_{{\rm L}q}-\mathcal{C}_{qr}\big)
	-\mu_{\rm i}\big(\partial_{\theta}f_{{\rm L}q}-\mathcal{C}_{q\theta}\big)
	\\
	&&-\mu_{\rm i}\sqrt{1-\mu_{\rm i}^2}\left(\cos\bar{\phi}_{\rm i}\partial_{\mu_{\rm i}}
	+\frac{\mu_{\rm i}\sin\bar{\phi}_{\rm i}}{1-\mu_{\rm i}^2}\partial_{\bar{\phi}_{\rm i}}
	\right)f_{{\rm L}q}
	\bigg]\,.
	\end{eqnarray} 	
One may implement the expressions for $\mathcal{L}^{\mu}_q$ above to evaluate each component of $T^{\mu\nu}$ from Equation~(\ref{J_and_T}) with the input of $f_{\rm L}$. Note that we only have to input the leading-order $O(\hbar^0)$ contributions of $\mathcal{C}^{\mu}_q$ into Equation~(\ref{Lf_inetrial}).

\subsection{Details of Collisions}
\label{sec_collision}
As mentioned previously, the collision terms depend on the underlying microscopic theory for the neutrino-matter interaction. To write down a more explicit expression of collisions and make a comparison with the classical collision terms widely applied in radiation hydrodynamics in core-collapse supernovae such as those in \cite{Bruenn:1985en, Reddy:1997yr}, we consider the weak interaction between neutrinos and nucleons as the matter sector. For simplicity and concreteness, we focus on the neutrino absorption on nucleons $(\nu_{\rm L}+{\rm n} \rightleftharpoons {\rm e}_{\rm L}+{\rm p})$ and elastic neutrino-nucleon scattering $(\nu_{\rm L}+{\rm N}\rightleftharpoons \nu_{\rm L}+{\rm N}$, where ${\rm N}={\rm n,p}$) based on the four-Fermi theory of the weak interaction. In addition to taking neutrinos as chiral fermions, we will assume that electrons are ultrarelativistic and treat them as approximate chiral fermions, which also incorporate quantum corrections due to the chirality. In the following, we will write down generic forms of lesser/greater self-energies and emission/absorption rates for each process above. 

To simplify the expressions mostly concerned with the quantum field theory calculations, we will take $c = 1$ below and in Appendices~\ref{app_four-Fermi_theory}-\ref{app_conserv_eq}, except for the final results in Equations~(\ref{Gamma_decomp})--(\ref{Gamma_omega}).

\subsubsection{Neutrino Absorption on Nucleons}
As the first example, we consider the neutrino absorption on nucleons
\begin{eqnarray}
\label{inelastic_s}
\nu^{\rm e}_{\rm L}(q)+{\rm n}(k)\rightleftharpoons {\rm e}_{\rm L}(q')+{\rm p}(k'),
\end{eqnarray}
where $q^{\mu}$ and $q'^{\mu}$ $(k^{\mu}$ and $k'^{\mu}$) correspond to the four-momenta of incoming or outgoing leptons (nucleons), respectively.
This process is described by the four-Fermi theory of the weak interaction, expressed as the current-current interactions,
\begin{eqnarray}
\mathcal{L}^{\rm cc}_{\rm int}=\frac{{G}_{\rm F}}{\sqrt{2}}(j_{\ell}^{-})_{\mu} (j_{\rm N}^{+})^{\mu}+{\rm h.c.}\,,
\end{eqnarray}
where $G_{\rm F}$ is the Fermi constant and 
$(j_{\ell}^{-})_{\mu} = \bar{\psi}_{\rm e} \gamma_{\mu}(1-\gamma^5)\psi_{\nu}$ 
and $(j_{\rm N}^{+})^{\mu} = \bar{\psi}_{{\rm p}}\gamma^{\mu}(g_{\rm V}-g_{\rm A}\gamma^5)\psi_{\rm n}$ are the lepton and nucleon charged currents, respectively, 
with $g_{\rm V}=1$ and $g_{\rm A} \approx 1.27$; for recent calculations of $g_{\rm A}$ in lattice QCD, see \cite{Chang:2018uxx} and references therein.

By a standard calculation, one finds the self-energies for this neutrino absorption process (see Appendix~\ref{app_four-Fermi_theory} for the details),
\begin{eqnarray}\label{Sigma_ab}
	\Sigma^{(\text{ab})\lessgtr}_{q\mu}&=&\int_p {\Pi}^{({\rm np})\lessgtr}_{p,\mu\nu}\mathcal{L}^{({\rm e})\lessgtr\nu}_{q-p},
\end{eqnarray} 
where
\begin{eqnarray}
{\Pi}^{({\rm np})\lessgtr}_{p,\mu\nu} = 8{G}_{\rm F}^2\int_k \left(g_{+}^2k_{\mu}k'_{\nu}+g_{-}^2 k'_{\mu}k_{\nu}-g_{+}g_{-}M_{\rm n}M_{\rm p}\eta_{\mu\nu}\right)
(2\pi)^2\delta(k^2-M_{\rm n}^2)\delta(k'^2-M_{\rm p}^2)f^{({\rm n})\gtrless}_{k}f^{({\rm p})\lessgtr}_{k'}\bigg|_{k'=p+k}\,,
\end{eqnarray}
with $p^{\mu} = (k'-k)^{\mu} = (q-q')^{\mu}$ being the four-momentum transfer in scattering, $M_{\rm n,p}$ the masses of neutrons/protons, $f^{({\rm N})<}$ and $f^{({\rm N})>}$ (${\rm N}={\rm n,p}$) the distribution functions of incoming and outgoing nucleons, respectively, and $g_{\pm} \equiv g_{\rm V}\pm g_{\rm A}$. 
In general, the Wigner functions for left-handed electrons here also incorporate quantum corrections, which have to be solved from another quantum transport equation for chiral fermions. More precisely, we shall take
\begin{eqnarray}
\label{L_e}
\mathcal{L}^{({\rm e})\lessgtr\mu}_q=2\pi\Big[\delta(q^2)\big(q^{\mu}-\hbar S^{\mu\nu}_{(\xi)q}\Delta_{q\nu}\big)-\hbar \tilde{F}^{\mu\nu}q_{\nu}\delta'(q^2)\Big]f^{({\rm e})\lessgtr}_{{\rm L}q}\,,
\end{eqnarray}
where we retain the electromagnetic fields coupled to electrons and $f_{\rm L}^{({\rm e})\lessgtr}$ denote the distribution functions of left-handed incoming/outgoing electrons with $n^{\mu}=\xi^{\mu}$. 
Generically, one has to solve for $f_{\rm L}^{({\rm e})\lessgtr}$ from a coupled CKT governing the dynamics of electrons. Then, we have
\begin{eqnarray}
\Sigma^{(\text{ab})\lessgtr}_{q\mu}=\int_p{\Pi}^{({\rm np})\lessgtr}_{p,\mu\nu}
\bigg[\delta(q'^2)\Big(q'^{\nu}-\hbar\frac{\epsilon^{\nu\rho\alpha\beta} q'_{\alpha}\xi_{\beta}}{2q'\cdot \xi}\Delta_{q' \rho}\Big)
-\hbar\tilde{F}^{\nu \rho}q'_{\rho}\delta'(q'^2)
\bigg]f_{{\rm L}q'}^{({\rm e})\lessgtr}\Bigg|_{q'=q-p}\,,
\end{eqnarray}
and accordingly,
\begin{eqnarray}\label{qsigma_ab}\nonumber
\Gamma^{(\text{ab})\lessgtr}_{(\xi)q}&=&
\int_p q^{\mu} {\Pi}^{({\rm np})\lessgtr}_{p,\mu\nu}
\Bigg\{ \bigg[\delta(q'^2) \Big(q'^{\nu}-\hbar\frac{\epsilon^{\nu\rho\alpha\beta}q'_{\alpha}\xi_{\beta}}{2q'\cdot \xi}\Delta_{q' \rho}\Big)
-\hbar\tilde{F}^{\nu\rho}q'_{\rho}\delta'(q'^2) \bigg]f_{{\rm L}q'}^{({\rm e})\lessgtr}
\\
&&+\hbar\frac{\epsilon^{\mu\rho\alpha\beta}q_{\alpha}\xi_{\beta}}{2q\cdot \xi}D_{q\rho}\int_p {\Pi}^{({\rm np})\lessgtr}_{p,\mu \nu} \delta(q'^2)q'^{\nu}
f_{{\rm L}q'}^{({\rm e})\lessgtr}\Bigg\}_{\! q'=q-p} \,.
\end{eqnarray}

Regarding the effects of electromagnetic fields, one can further focus on an external magnetic field such that $\tilde{F}^{\mu\nu}=B^{\mu}\xi^{\nu}-B^{\nu}\xi^{\mu}$ and $F^{\mu\nu}=-\epsilon^{\mu\nu\alpha\beta}B_{\alpha}\xi_{\beta}$. In this case, Equation (\ref{qsigma_ab}) becomes
\begin{eqnarray}\label{qsigma_ab_B}
\Gamma^{(\text{ab})\lessgtr}_{(\xi)q}
&=&\int_p q^{\mu} {\Pi}^{({\rm np})\lessgtr}_{p, \mu \nu} \Bigg\{ \delta(q'^2) \bigg[\Big(q'^{\nu}-\hbar\frac{\epsilon^{\nu\rho\alpha\beta}q'_{\alpha}\xi_{\beta}}{2q'\cdot\xi}D_{q'\rho}\Big)+\frac{\hbar}{2q'\cdot\xi}
\Big(B\cdot q'\partial_{q'^{\nu}_{\perp}}-B_{\nu}q'_{\rho}\partial_{q'^{\rho}_{\perp}}\Big)\bigg]f_{{\rm L}q'}^{({\rm e})\lessgtr}
\nonumber \\
&&-\hbar\Big(B_{\nu}q'\cdot\xi-\xi_{\nu}B\cdot q'\Big)\delta'(q'^2)
f_{{\rm L}q'}^{({\rm e})\lessgtr}
+\hbar\frac{\epsilon^{\mu\rho\alpha\beta}q_{\alpha}\xi_{\beta}}{2q\cdot \xi}D_{q\rho}\int_p {\Pi}^{({\rm np})\lessgtr}_{p,\mu \nu} \delta(q'^2) q'^{\nu}
f_{{\rm L}q'}^{({\rm e})\lessgtr} \Bigg\}_{\! q'=q-p} \,,
\end{eqnarray}
where $V_{\perp}^{\mu}=(g^{\mu\nu}-\xi^{\mu}\xi^{\nu})V_{\nu}$ represents the component perpendicular to the frame vector $\xi$ for an arbitrary vector $V$. Despite the complexity of Equation (\ref{qsigma_ab_B}), we can write down the following structure based on the symmetry of the system:
\begin{eqnarray}\label{structure_Gammaab}
\Gamma^{(\text{ab})\lessgtr}_{(\xi)q}= \Gamma^{(0)\lessgtr}_{q}
+\hbar\epsilon^{\mu\nu\alpha\beta}q_{\mu}\xi_{\nu}\Big(\Gamma^{(1)\lessgtr}_{q}\partial_{\alpha}U_{\beta}+U_{\alpha}\partial_{\beta}\Gamma^{(2)\lessgtr}_{q}\Big)
+\hbar\Big(\Gamma^{(3)\lessgtr}_{q}q\cdot B+\Gamma^{(4)\lessgtr}_{q}U\cdot B\Big),
\end{eqnarray}
where $\Gamma^{(0)\lessgtr}_{q}$ is the classical collision term, $\Gamma^{(k)\lessgtr}_{q}$ ($k=1,\cdots,4$) are the quantum corrections related to the chirality of fermions, and $U_{\mu}(x)$ is a vector characterizing local properties of the matter. The detailed structure of $U^{\mu}$ and the coefficients $\Gamma^{(k)\lessgtr}_{q}$ ($k=0,1,\cdots,4$) depend on the nucleon and electron distribution functions. For instance, assuming that the matter is in local thermal equilibrium, $U_{\mu}$ corresponds to the local fluid four-velocity and $\partial_{\beta}\Gamma^{(2)\lessgtr}_{q}$ contains the spatial derivatives of local temperature or chemical potentials.

\subsubsection{Elastic Neutrino-Nucleon Scattering}
We next consider the elastic neutrino-nucleon scattering
\begin{eqnarray}
\label{elastic_s}
\nu^{\ell}_{\rm L}(q)+{\rm N}(k)\rightleftharpoons \nu^{\ell}_{\rm L}(q')+{\rm N}(k').
\end{eqnarray}
This is described by the current-current interactions of the form
\begin{eqnarray}
\mathcal{L}^{\rm nc}_{\rm int}=\frac{G_{\rm F}}{\sqrt{2}}(j_{\nu})_{\mu} (j_{\rm N})^{\mu}+{\rm h.c.}\,,
\end{eqnarray}
where $(j_{\nu})_{\mu} = \bar{\psi}_{\nu}\gamma_{\mu}(1-\gamma^5)\psi_{\nu}$ 
and $(j_{\rm N})^{\mu} = \frac{1}{2}\bar{\psi}_{\rm N}\gamma^{\mu}(c_{\rm V}-c_{\rm A}\gamma^5)\psi_{\rm N}$ are the lepton and nucleon neutral currents, respectively.
Here $c_{\rm V}= -1$ and $c_{\rm A} = - g_{\rm A}$ for ${\rm N} = {\rm n}$, 
and $c_{\rm V}= 1-4 \sin^2 \theta_{\rm W}$ and $c_{\rm A} = g_{\rm A}$ for ${\rm N} = {\rm p}$, where $\theta_{\rm W}$ is the Weinberg angle; see, e.g., \cite{Reddy:1997yr}.

The self-energies for this process are given by (see Appendix~\ref{app_four-Fermi_theory})
\begin{eqnarray}\nonumber
\Sigma^{(\text{el})\lessgtr}_{q\mu}&=&\int_p {\Pi}^{({\rm NN})\lessgtr}_{p,\mu\lambda}\mathcal{L}^{(\nu)\lessgtr\lambda}_{q-p}
\\
&=&\int_p \bigg[\delta(q'^2){\Pi}^{({\rm NN})\lessgtr}_{p,\mu\lambda}\Big(q'^{\lambda}-\hbar\frac{\epsilon^{\lambda\rho\alpha\beta}q'_{\alpha}\xi_{\beta}}{2q'\cdot \xi}D_{q'\rho}\Big)f_{{\rm L}q'}^{(\nu)\lessgtr}\bigg]_{q'=q-p}\,,
\end{eqnarray}
where 
\begin{eqnarray}
{\Pi}^{({\rm NN})\lessgtr}_{p,\mu\nu}
=8G_{\rm F}^2\int_k\left(c_{+}^2k_{\mu}k'_{\nu} +c_{-}^2k'_{\mu}k_{\nu}-c_{+}c_{-}M_{\rm N}^2\eta_{\mu\nu}\right)
(2\pi)^2\delta(k^2-M_{\rm N}^2)\delta(k'^2-M_{\rm N}^2)f^{({\rm N})\gtrless}_{k}f^{({\rm N})\lessgtr}_{k'}\bigg|_{k'=p+k}\,,
\end{eqnarray}
with $c_{\pm} \equiv (c_{\rm V}\pm c_{\rm A})/2$, and 
\begin{eqnarray}
\label{L_nu}
\mathcal{L}^{({\nu})\lessgtr\mu}_q=2\pi\Big[\delta(q^2)\big(q^{\mu}-\hbar S^{\mu\nu}_{(\xi)q}D_{q\nu}\big)\Big]f^{(\nu)\lessgtr}_{{\rm L}q}\,.
\end{eqnarray}
Consequently, we find
\begin{eqnarray}
\Gamma^{(\text{el})\lessgtr}_{(\xi)q}
=\int_p q^{\mu} {\Pi}^{({\rm NN})\lessgtr}_{p, \mu \lambda}\bigg[\delta(q'^2)\Big(q'^{\lambda}-\hbar\frac{\epsilon^{\lambda\rho\alpha\beta}q'_{\alpha}\xi_{\beta}}{2q'\cdot\xi}D_{q'\rho}\Big)f_{{\rm L}q'}^{(\nu)\lessgtr}+\hbar\frac{\epsilon^{\mu\rho\alpha\beta}q_{\alpha}\xi_{\beta}}{2q\cdot \xi}D_{q\rho}\int_p {\Pi}^{({\rm NN})\lessgtr}_{p,\mu \lambda} \delta(q'^2) q'^{\lambda} 
f_{{\rm L}q'}^{(\nu)\lessgtr}\bigg]_{q'=q-p}\,,
\end{eqnarray}
which is similar to the form in Equation~(\ref{qsigma_ab_B}) without background fields. Nevertheless, because the neutrino distribution functions are involved in the integrand, one has to make further approximations to simplify the nonlinear terms in neutrino distribution functions. For example, the so-called isoenergetic approximation by assuming zero-energy transfer may be used \citep{Bruenn:1985en}. In this approximation, one finds ${\Pi}^{({\rm NN})<\nu}_{p,\lambda}={\Pi}^{({\rm NN})>\nu}_{p,\lambda}\equiv {\Pi}^{({\rm NN})\nu}_{p,\lambda}$, as can be easily shown in thermal equilibrium with detailed balance. Therefore, one can linearize the collision term in the kinetic theory with the isoenergetic approximation as
\begin{eqnarray}
(1-f^{(\nu)}_{{\rm L}q})\Gamma^{(\text{el})<}_{(\xi)q}-f^{(\nu)}_{{\rm L}q}\Gamma^{(\text{el})>}_{(\xi)q}
&=&\int_p q^{\mu} {\Pi}^{({\rm NN})}_{p,\mu\lambda}\Bigg\{\delta(q'^2)\bigg[q'^{\lambda}\Big(f^{(\nu)}_{{\rm L}q'}-f^{(\nu)}_{{\rm L}q}\Big)-\hbar\frac{\epsilon^{\lambda\rho\alpha\beta}q'_{\alpha}\xi_{\beta}}{2q'\cdot\xi}D_{q'\rho}f^{(\nu)}_{{\rm L}q'}\bigg]
\nonumber \\
&&
+\hbar\frac{\epsilon^{\mu\rho\alpha\beta}q_{\alpha}\xi_{\beta}}{2q\cdot \xi}\bigg(D_{q\rho}\int_p {\Pi}^{({\rm NN})}_{p,\mu \lambda} \delta(q'^2) q'^{\lambda}
f^{(\nu)}_{{\rm L}q'}-f^{(\nu)}_{{\rm L}q}D_{q\rho}\int_p {\Pi}^{({\rm NN})}_{p,\mu \lambda}\delta(q'^2) q'^{\lambda} \bigg) \Bigg\}_{\! q'=q-p}\,.
\end{eqnarray}

When neglecting the quantum corrections involving the Levi-Civita tensor $\epsilon^{\mu \nu \alpha \beta}$ or the magnetic field $B^{\mu}$, our collision terms for both processes reduce to those presented in \cite{Reddy:1997yr}.

\subsection{Collisions with Matter in Equilibrium}
In the case of core-collapse supernovae, we can assume that the matter sector consisting of nucleons and electrons is in local thermal equilibrium and can be described by hydrodynamics, since the typical length scale of interest is much larger than their mean free paths. This allows us to derive an analytic form of the collision term by employing proper approximations, e.g., for the neutrino absorption process. The computations of the quantum corrections in other processes will be reported elsewhere.

First, we assume that the fluid velocity is sufficiently small such that $u^{\mu}=\gamma(1,{\bm v})\approx (1,{\bm 0})=\xi^{\mu}$, and hence $\tilde{F}^{\mu\nu}\approx B^{\mu}u^{\nu}-B^{\nu}u^{\mu}$.%
\footnote{In general, in the presence of a magnetic field in the inertial frame, there will be a nonvanishing electric field in the comoving frame with respect to the fluid four-velocity. In this case, the system is inevitably driven away from equilibrium. For simplicity, we make the nonrelativistic fluid approximation to ignore such an electric field.}
Second, as we are here interested in the quantum corrections due to the vorticity $\omega^{\mu}\equiv\frac{1}{2}\epsilon^{\mu\nu\alpha\beta}u_{\nu}\nabla_{\alpha}u_{\beta}$ and the magnetic field $B^{\mu}$, we will ignore the viscous corrections and the gradients of the temperature and chemical potentials. Under such assumptions, we can ignore the terms $\nabla_{\{\mu}u_{\nu\}}$ and $u^{\nu} \nabla_{\nu} u^{\mu}$,%
\footnote{Note that we will ignore the terms $\nabla_{\{\mu}u_{\nu\}}$ and $u^{\nu} \nabla_{\nu} u^{\mu}$ only for the computations of the quantum corrections in the collisions here, but such terms should be included to describe the hydrodynamic evolutions of the system.}
and as a result, we have $\nabla_{\mu}u_{\nu}\approx \nabla_{[\mu}u_{\nu]}\approx-\epsilon_{\mu\nu\alpha\beta}\omega^{\alpha}u^{\beta}$.
According to \cite{Hidaka:2017auj}, the lesser/greater propagators for left-handed thermal electrons can then be written as
\begin{eqnarray}
	\mathcal{\bar L}^{({\rm e})<\mu}_q&=&2\pi\bigg[\delta(q^2)\Big(q^{\mu}f^{({\rm e})}_{0,q}-\frac{\hbar\beta}{2}\big(\omega^{\mu}q\cdot u-u^{\mu}q\cdot \omega\big)f^{({\rm e})}_{0,q}(1-f^{({\rm e})}_{0,q})\Big)
	-\hbar\big(B^{\mu}q\cdot u-u^{\mu}q\cdot B\big)\delta'(q^2)f^{({\rm e})}_{0,q}\bigg], 
	\nonumber \\
	\mathcal{\bar L}^{({\rm e})>\mu}_q&=&2\pi\bigg[\delta(q^2)\Big(q^{\mu}(1-f^{({\rm e})}_{0,q})+\frac{\hbar\beta}{2}\big(\omega^{\mu}q\cdot u-u^{\mu}q\cdot \omega\big)f^{({\rm e})}_{0,q}(1-f^{({\rm e})}_{0,q})\Big)
	-\hbar\big(B^{\mu}q\cdot u-u^{\mu}q\cdot B\big)\delta'(q^2)(1-f^{({\rm e})}_{0,q})\bigg], 
	\nonumber \\
\end{eqnarray}
where
\begin{eqnarray}
	f^{(i)}_{0,q}=\frac{1}{{\rm e}^{\beta (q\cdot u-\mu_i)}+1}\,, \quad (i={\rm n}, {\rm p}, {\rm e})
\end{eqnarray}
represent the Fermi-Dirac distribution functions with $\beta=1/(k_{\rm B}T)$, with $T$, $\mu_i$, and $k_{\rm B}$ being temperature, chemical potentials for $i={\rm n}, {\rm p}, {\rm e}$, and Boltzmann constant, respectively. Here and below, $\mathcal{\bar O}$ stands for a quantity $\mathcal{O}$ in local thermal equilibrium.

Now the self-energies in Equation~(\ref{Sigma_ab}) become
\begin{eqnarray}\label{self_E_eq}
\Sigma^{\gtrless}_{\mu}=\bar \Sigma^{(0)\gtrless}_{\mu}+ \hbar \bar \Sigma^{(\omega)\gtrless}_{\mu}+ \hbar \bar \Sigma^{(B)\gtrless}_{\mu}\,,
\end{eqnarray} 
where
\begin{eqnarray}\nonumber\label{Sigma0eq}
\bar \Sigma^{(0)\gtrless}_{\mu}&=& 8{G}_{\rm F}^2\int_p\int_k\Big(g_{+}^2k_{\mu}k'_{\nu} 
+g_{-}^2 k'_{\mu}k_{\nu}-g_{+}g_{-}M_{\rm n}M_{\rm p}\eta_{\mu\nu} \Big)
q'^{\nu}(2\pi)^3\delta(q'^2)\delta(k^2-M_{\rm n}^2)\delta(k'^2-M_{\rm p}^2)
\\
&& \times
f^{({\rm n})\lessgtr}_{0,k} f^{({\rm p})\gtrless}_{0,k'}f^{({\rm e})\gtrless}_{0,q'}\Big|_{q'=q-p, \ k'=p+k}\,,
\\
\nonumber\label{Sigma_omega_eq}
\bar \Sigma^{(\omega)\gtrless}_{\mu}&=&\mp 4\beta {G}_{\rm F}^2\int_p\int_{k}\Big(g_{+}^2k_{\mu}\big[ (\omega \cdot k')(q' \cdot u) - (\omega \cdot q')(k' \cdot u) \big]
+g_{-}^2 k'_{\mu}\big[(\omega \cdot k)(q' \cdot u) - (\omega \cdot q')(k\cdot u) \big]
\\\nonumber
&&
-g_{+}g_{-}M_{\rm n}M_{\rm p}\big[\omega_{\mu} (q' \cdot u) - u_{\mu} (q' \cdot\omega) \big]
\Big) (2\pi)^3\delta(q'^2)\delta(k^2-M_{\rm n}^2)\delta(k'^2-M_{\rm p}^2)
\\
&&\times f^{({\rm n})\lessgtr}_{0,k} f^{({\rm p})\gtrless}_{0,k'} f^{({\rm e})}_{0,q'}(1-f^{({\rm e})}_{0,q'})\Big|_{q'=q-p, \ k'=p+k}\,, 
\\
\nonumber\label{Sigma_B_eq}
\bar \Sigma^{(B)\gtrless}_{\mu}&=&-8{G}_{\rm F}^2 \int_p\int_k\Big(g_{+}^2k_{\mu}\big[(B \cdot k')(q'\cdot u) - (B \cdot q')(k'\cdot u) \big]
+g_{-}^2 k'_{\mu}\big[(B \cdot k)(q' \cdot u)-(B \cdot q')(k\cdot u)\big]
\\\nonumber
&&
-g_{+}g_{-}M_{\rm n}M_{\rm p}\big[B_{\mu} (q'\cdot u) - u_{\mu} (q'\cdot B) \big]
\Big) (2\pi)^3\delta'(q'^2)\delta(k^2-M_{\rm n}^2)\delta(k'^2-M_{\rm p}^2)
\\
&&\times 
f^{({\rm n})\lessgtr}_{0,k}f^{({\rm p})\gtrless}_{0,k'}f^{({\rm e})\gtrless}_{0,q'}\Big|_{q'=q-p, \ k'=p+k}\,.
\end{eqnarray}
Here we drop the contributions of antiparticles for simplicity and work in Minkowski spacetime for the integrals. 
Accordingly, the absorption and radiation rates in the collision term take the form
\begin{eqnarray}
\Gamma^{\lessgtr}_{(\xi) q}=q\cdot \bar \Sigma^{(0)\lessgtr}+ \hbar \big(q\cdot \bar \Sigma^{(\omega)\lessgtr}+q\cdot \bar \Sigma^{(B)\lessgtr}\big)
-\hbar S^{\mu\nu}_{(\xi)q}D^{({\rm i})}_{q\mu} \bar \Sigma^{(0)\lessgtr}_{\nu}\,.
\end{eqnarray}

In the following, we will set $M_{\rm n} \approx M_{\rm p} \approx M$ and adopt the nonrelativistic approximation for nucleons. 
We will also use the ``quasi-isoenergetic" approximation that allows for the energy transfer up to $O(1/M)$ (see Appendix \ref{app_self_energies}).
One then finds
\begin{eqnarray}\nonumber\label{Sigma_mu_0B}
\bar \Sigma^{(0)>}_{\mu}+\hbar \bar \Sigma^{(B)>}_{\mu}&\approx&  \frac{1}{{\pi}}\big(g_{\rm V}^2+3g_{\rm A}^2\big){G}_{\rm F}^2
u_{\mu}|{\bm q}|^2(1-f^{({\rm e})}_{0,q})\left(1-\frac{3|{\bm q}|}{M}-\frac{\hbar B_{\rm L}}{2M|{\bm q}|}\right)\frac{n_{\rm n}-n_{{\rm p}}}{1-{\rm e}^{\beta(\mu_{{\rm p}}-\mu_{\rm n})}}\,,
\\
\bar \Sigma^{(0)<}_{\mu}+\hbar \bar \Sigma^{(B)<}_{\mu}&\approx&  \frac{1}{\pi}\big(g_{\rm V}^2+3g_{\rm A}^2\big){G}_{\rm F}^2
u_{\mu}|{\bm q}|^2f^{({\rm e})}_{0,q}\left(1-\frac{3|{\bm q}|}{M}-\frac{\hbar B_{\rm L}}{2M|{\bm q}|}\right)\frac{n_{{\rm p}}-n_{\rm n}}{1-{\rm e}^{\beta(\mu_{\rm n}-\mu_{{\rm p}})}}\,,
\end{eqnarray}
and
\begin{eqnarray}\nonumber\label{Sigma_mu_omega}
	\hbar\Big(q\cdot \bar \Sigma^{(\omega)>}+ S^{\mu\nu}_{(n)}D_{\mu} \bar \Sigma^{(0)>}_{\nu}\Big)
	&\approx& \frac{\hbar}{2\pi} \big(g_{\rm V}^2+3g_{\rm A}^2\big){G}_{\rm F}^2 |{\bm q}|(q\cdot \omega)(1-f^{({\rm e})}_{0,q})(2+\beta q_0f^{({\rm e})}_{0,q})
	\frac{n_{\rm n}-n_{{\rm p}}}{1-{\rm e}^{\beta(\mu_{{\rm p}}-\mu_{\rm n})}}\,,
	\\
	\hbar\Big(q\cdot \bar \Sigma^{(\omega)<}+ S^{\mu\nu}_{(n)}D_{\mu} \bar \Sigma^{(0)<}_{\nu}\Big)
	&\approx& \frac{\hbar}{2\pi} \big(g_{\rm V}^2+3g_{\rm A}^2\big){G}_{\rm F}^2 |{\bm q}|(q\cdot \omega)f^{({\rm e})}_{0,q}\big(2-\beta q_0(1-f^{({\rm e})}_{0,q})\big)
	\frac{n_{{\rm p}}-n_{\rm n}}{1-{\rm e}^{\beta(\mu_{\rm n}-\mu_{{\rm p}})}}\,,
\end{eqnarray}
where $|{\bm q}|\equiv \sqrt{-q_{\perp}^2}$ and $n_{{\rm n}/{\rm p}}\equiv \int \frac{{\rm d}^3 {\bm k}}{(2\pi)^{3}} f^{({\rm n}/{\rm p})}_{0,k}$ are neutron/proton densities, 
and we decomposed the magnetic field into the longitudinal and transverse components with respect to the momentum $q_{\perp}^{\mu}$ as
\begin{eqnarray}
\label{B_decomp}
B^{\mu}=\frac{q_{\perp}^{\mu}}{|{\bm q}|} B_{\rm L} + B_{\rm T}^{\mu}\,, \qquad q_{\perp} \cdot B_{\rm T} = 0\,.
\end{eqnarray} 
Note that while the quantum corrections due to magnetic fields are suppressed in the $M \rightarrow \infty$ limit, those corrections due to the fluid vorticity persist even in this limit. 

Assembling all pieces together, taking $|{\bm q}|\approx E_{\rm i}$ and restoring $\hbar$ and $c$, Equation~(\ref{structure_Gammaab}) in thermal equilibrium reduces to (now $U^{\mu}=u^{\mu}$)
\begin{eqnarray}
	\bar \Gamma^{\lessgtr}_{(\xi)q}\approx \bar \Gamma^{(0)\lessgtr}_{q}
	+\hbar \bar \Gamma^{(\omega)\lessgtr}_{q} (q\cdot \omega)
	+\hbar \bar \Gamma^{(B)\lessgtr}_{q} (q\cdot B),
	\label{Gamma_decomp}
\end{eqnarray}
where
\begin{eqnarray}\nonumber
	\bar \Gamma^{(0)>}_{q}&\approx& \frac{1}{\pi \hbar^4 c^4} \big(g_{\rm V}^2+3g_{\rm A}^2\big){G}_{\rm F}^2 E_{\rm i}^3(1-f^{({\rm e})}_{0,q})\left(1-\frac{3E_{\rm i}}{M c^2}\right)	\frac{n_{{\rm p}}-n_{\rm n}}{1-{\rm e}^{\beta(\mu_{\rm n}-\mu_{{\rm p}})}}=2M E_{\rm i}^2\left(1-\frac{E_{\rm i}}{3M c^2}\right) \bar \Gamma^{(B)>}_{q}\,,
	\\
	\bar \Gamma^{(0)<}_{q}&\approx& \frac{1}{\pi \hbar^4 c^4} \big(g_{\rm V}^2+3g_{\rm A}^2\big){G}_{\rm F}^2 E_{\rm i}^3f^{({\rm e})}_{0,q}\left(1-\frac{3E_{\rm i}}{M c^2}\right)	\frac{n_{{\rm p}}-n_{\rm n}}{1-{\rm e}^{\beta(\mu_{\rm n}-\mu_{{\rm p}})}}=2M E_{\rm i}^2\left(1-\frac{E_{\rm i}}{3M c^2}\right) \bar \Gamma^{(B)<}_{q}\,,
	\label{Gamma_B}
\end{eqnarray}
and
\begin{eqnarray}\nonumber
	\bar \Gamma^{(\omega)>}_{q}&\approx & \frac{1}{2\pi \hbar^4 c^4} \big(g_{\rm V}^2+3g_{\rm A}^2\big){G}_{\rm F}^2 E_{\rm i}(1-f^{({\rm e})}_{0,q})(2+\beta E_{\rm i} f^{({\rm e})}_{0,q})
		\frac{n_{{\rm p}}-n_{\rm n}}{1-{\rm e}^{\beta(\mu_{\rm n}-\mu_{{\rm p}})}}\,,
	\\
	\bar \Gamma^{(\omega)<}_{q}&\approx & \frac{1}{2\pi \hbar^4 c^4} \big(g_{\rm V}^2+3g_{\rm A}^2\big){G}_{\rm F}^2 E_{\rm i}f^{({\rm e})}_{0,q}\big(2-\beta E_{\rm i}(1-f^{({\rm e})}_{0,q})\big)
	\frac{n_{{\rm p}}-n_{\rm n}}{1-{\rm e}^{\beta(\mu_{\rm n}-\mu_{{\rm p}})}}\,.
	\label{Gamma_omega}
\end{eqnarray}
Consequently, the emission and absorption rates are obtained as $R_{\rm emis} = c \bar \Gamma^{<}_{q}/E_{\rm i}$ and $R_{\rm abs} = c\Gamma^{>}_{q}/E_{\rm i}$, respectively.

Finally, we discuss the physical consequences of these quantum corrections. First of all, both the $q \cdot \omega$ and $q \cdot B$ terms break the spherical symmetry and axisymmetry of the system, as we already argued in a generic frame. Note also that these terms break the parity symmetry, which is a feature specific to the parity-violating weak interaction. Moreover, an important feature of the $q \cdot \omega$ and $q \cdot B$ terms is that, for the neutrinos propagating collinear to the flow of matter, they can give leading-order contributions to the so-called kinetic helicity ${\bm v} \cdot {\bm \omega}$ and cross helicity ${\bm v} \cdot {\bm B}$ of the matter, respectively, where ${\bm \omega} \equiv \frac{1}{2} {\bm \nabla} \times {\bm v}$.%
\footnote{Note that, although the higher-order terms in $|{\bm v}|$ are dropped here as subleading corrections in collisions, the contributions of these collision terms to the helicity are not suppressed by $|{\bm v}|$. In fact, by decomposing $q^{\mu}$ as $q^{\mu}=(q\cdot v)v^{\mu}/|{\bm v}|^2+q_{\rm T}^{\mu}$ and $v\cdot q_{\rm T}=0$ with $v^{\mu}\equiv(0,{\bm v})$, we have $q\cdot \omega= (q\cdot v) (v\cdot \omega)/ |{\bm v}|^2+q_{\rm T}\cdot \omega$ and $q\cdot B= (q\cdot v) (v\cdot B)/|{\bm v}|^2+q_{\rm T}\cdot B$, which show that their contributions to the helicity are not suppressed by $|{\bm v}|$.} The mechanism that chiral effects of neutrinos, combined with the neutrino-matter interaction, can generate the kinetic helicity and cross helicity of the matter was previously shown in \citet{Yamamoto:2015gzz} in the hydrodynamic regime of neutrinos. The new collision terms above provide its generalization to the case away from equilibrium, where hydrodynamics for neutrinos is not necessarily applicable. The presence of the kinetic helicity of the matter further induces magnetic helicity by the helical plasma instability \citep{Yamamoto:2015gzz}, and as a result, it gives the tendency toward the inverse energy cascade \citep{Masada:2018swb}, which would be favorable for the supernova explosion.

In the presence of a background magnetic field ${\bm B}_{\rm ex}$ and/or a global rotation of the system characterized by the angular velocity ${\bm \Omega}$, we have the collision terms of the form ${\bm q}\cdot {\bm B}_{\rm ex}$ and/or ${\bm q} \cdot {\bm \Omega}$. These collision terms lead to the asymmetric neutrino emission with respect to the directions of ${\bm B}_{\rm ex}$ and ${\bm \Omega}$, respectively, which may contribute to the pulsar kick.

\section{Summary and outlook}
\label{sec_summary}
In this work, we have constructed the chiral radiation transport equation for left-handed neutrinos with the quantum corrections due to their chirality, mainly in the inertial frame. We have also shown the expression of the radiative energy-momentum tensor with quantum corrections via the Wigner functions. In particular, we derive the analytic forms of the emission and absorption rates including the quantum corrections for the neutrino absorption process. The formalism of neutrino chiral radiation hydrodynamics established in our work should be applied to perform numerical simulations for core-collapse supernova explosions and neutron star formation in future. 

In principle, one can also develop the same formalism in the comoving frame, while a different frame vector may be chosen for computational convenience. However, even for the ordinary 3D Boltzmann equation without quantum corrections, the free-streaming part involving fluid velocity in such a coordinate system is rather complicated \citep{1986Ap&SS.121..105M,Castor_09}. The generalization to further include quantum corrections seems to be technically difficult in that direction.

On the other hand, in order to further explore nonequilibrium chiral transport of electrons in supernovae, one may employ a kinetic theory with quantum corrections for massive fermions that has been developed more recently \citep{Weickgenannt:2019dks,Gao:2019znl,Hattori:2019ahi,Yang:2020hri,Liu:2020flb}. In particular, the quantum kinetic theory developed in \citet{Yang:2020hri} systematically includes collisional effects.

Finally, although we have focused on the chiral radiation transfer of neutrinos in this paper, our formulation here may also be extended to the radiative transfer of photons that incorporates the effects of their circular polarizations; see \citet{Yamamoto:2017uul,Huang:2018aly} for the CKT of photons. Such a formulation would be applicable to a variety of astrophysical systems involving photon radiation.

\acknowledgments
We thank K.~Mameda, Y.~Suwa, and T.~Takiwaki for useful conversations.
This work was supported by the Keio Institute of Pure and Applied Sciences (KiPAS) project at Keio University and MEXT-Supported Program for the Strategic Research Foundation at Private Universities, ``Topological Science'' (grant No.~S1511006). N.~Y. was supported by JSPS KAKENHI grant No.~19K03852.

\appendix
\section{Useful relations for coordinate transformation}
\label{app_relations}
	For a function $f(q)$ with the on-shell condition $q^2=0$, we have
	\begin{eqnarray}\label{df_1}
		{\rm d}f(q^r,q^{\theta},q^{\phi})=\left(\frac{\partial f}{\partial q^r}\right){\rm d}q^r+\left(\frac{\partial f}{\partial q^{\theta}}\right){\rm d}q^{\theta}+\left(\frac{\partial f}{\partial q^{\phi}}\right){\rm d}q^{\phi}\,.
	\end{eqnarray}	
	Considering $f(E,\mu,\bar{\phi})$, where we dropped the irrelevant spacetime dependence of $f$ for simplicity, we also find
	\begin{eqnarray}\nonumber\label{df_2}
		{\rm d}f&=&\left(\frac{\partial f}{\partial E}\right){\rm d}E+\left(\frac{\partial f}{\partial \mu}\right){\rm d}\mu
		+\left(\frac{\partial f}{\partial \bar{\phi}}\right){\rm d}\bar{\phi}
		\\\nonumber
		&=&\left(\mu \frac{\partial f}{\partial E} +\frac{1-\mu^2}{E} \frac{\partial f}{\partial \mu} \right){\rm e}^{\Lambda}{\rm d}q^r
		+R\sqrt{1-\mu^2}\Bigg[\left(\frac{\partial f}{\partial E} -\frac{\mu}{E} \frac{\partial f}{\partial \mu} \right)\cos\bar{\phi}
		-\frac{\sin\bar{\phi}}{E(1-\mu^2)} \frac{\partial f}{\partial\bar{\phi}} \Bigg]
		{\rm d}q^{\theta}
		\\
		&&+R\sin\theta\sqrt{1-\mu^2}\Bigg[\left(\frac{\partial f}{\partial E}-\frac{\mu}{E} \frac{\partial f}{\partial \mu}\right) \sin\bar{\phi}+\frac{\cos\bar{\phi}}{E(1-\mu^2)} \frac{\partial f}{\partial\bar{\phi}} \Bigg]
		{\rm d}q^{\phi}\,,
	\end{eqnarray}
	where we employed the relations
	\begin{eqnarray}
		{\rm d}E&=&\mu {\rm e}^{\Lambda}{\rm d}q^r+R\sqrt{1-\mu^2}\big(\cos\bar{\phi} {\rm d}q^{\theta}
		+\sin\theta\sin\bar{\phi} {\rm d}q^{\phi}
		\big)\,,
		\\
		{\rm d}\mu&=&\frac{{\rm e}^{\Lambda}}{E}(1-\mu^2){\rm d}q^r-\frac{\mu R}{E}\sqrt{1-\mu^2}\big(\cos\bar{\phi} {\rm d}q^{\theta}
		+\sin\theta\sin\bar{\phi} {\rm d}q^{\phi}
		\big)\,,
		\\
		{\rm d}\bar{\phi}&=&-\frac{R}{E\sqrt{1-\mu^2}}\big(\sin\bar{\phi} {\rm d}q^{\theta}-\sin\theta\cos\bar{\phi}{\rm d}q^{\phi}\big)\,,
	\end{eqnarray}
	which are obtained from
	\begin{eqnarray}\nonumber
		&&{\rm d}q^t={\rm e}^{-\Phi}{\rm d}E\,, \quad {\rm d}q^r={\rm e}^{-\Lambda}(\mu {\rm d}E+E{\rm d}\mu)\,,
		\quad
		{\rm d}q^{\theta}=\frac{\sqrt{1-\mu^2}}{R}\Big(\cos\bar{\phi}{\rm d}E-\frac{\mu E\cos\bar{\phi}}{1-\mu^2}{\rm d}\mu-E\sin\bar{\phi}{\rm d}\bar{\phi}\Big)\,,
		\\ 
		&&
		{\rm d}q^{\phi}=\frac{\sqrt{1-\mu^2}}{R\sin\theta}\Big(\sin\bar{\phi}{\rm d}E-\frac{\mu E\sin\bar{\phi}}{1-\mu^2}{\rm d}\mu+E\cos\bar{\phi}{\rm d}\bar{\phi}\Big)\,.
	\end{eqnarray}
	Comparing Equations~(\ref{df_1}) and (\ref{df_2}), we derive
	\begin{eqnarray}\nonumber\label{der_rel}
		\frac{\partial f}{\partial q^r}&=&\left(\mu \frac{\partial f}{\partial E} +\frac{1-\mu^2}{E} \frac{\partial f}{\partial \mu} \right){\rm e}^{\Lambda}\,,
		\\\nonumber
		\frac{\partial f}{\partial q^{\theta}}&=&R\sqrt{1-\mu^2}\Bigg[\left(\frac{\partial f}{\partial E}-\frac{\mu}{E}\frac{\partial f}{\partial \mu}\right)\cos\bar{\phi}
		-\frac{\sin\bar{\phi}}{E(1-\mu^2)}\frac{\partial f}{\partial\bar{\phi}}\Bigg]\,,
		\\
		\frac{\partial f}{\partial q^{\phi}}&=&R\sin\theta\sqrt{1-\mu^2}\Bigg[\left(\frac{\partial f}{\partial E}-\frac{\mu}{E}\frac{\partial f}{\partial \mu}\right)\sin\bar{\phi}+\frac{\cos\bar{\phi}}{E(1-\mu^2)} \frac{\partial f}{\partial\bar{\phi}} \Bigg]\,.
	\end{eqnarray}

\section{Perturbative solution of Wigner functions}
\label{app_Wigner}
\setcounter{equation}{0}
We shall show that Equation~(\ref{WF_L_full}) is the solution of Equation~(\ref{m_eq3}) up to $O(\hbar)$.
	Taking 
	\begin{eqnarray}
		\mathcal{L}^{\mu}=2\pi\Big[\delta(q^2)\big(q^{\mu}-\hbar c S^{\mu\nu}_{(n)}\mathcal{D}_{\nu}\big)-\hbar c \tilde{F}^{\mu\nu}q_{\nu}\delta'(q^2)\Big]f_{\rm L},
	\end{eqnarray}
	we find
	\begin{eqnarray}
		\hbar c \big(\mathcal{D}_{\mu}\mathcal{L}_{\nu}-\mathcal{D}_{\nu}\mathcal{L}_{\mu}\big)
		=2\pi \hbar c \Big[\delta(q^2)\big(2F_{\nu\mu}+2q_{[\nu}\mathcal{D}_{\mu]}\big)+4q^{\rho}F_{\rho[\mu}q_{\nu]}\delta'(q^2)\Big]f_{\rm L}+O(\hbar^2).
	\end{eqnarray}
	On the other hand, we have
	\begin{eqnarray}\nonumber
		2\epsilon_{\mu\nu\rho\sigma}q^{\rho}\mathcal{L}^{\sigma}
		&=&-4\pi \hbar c \epsilon_{\mu\nu\rho\sigma}q^{\rho}\bigg[\delta(q^2)\frac{\epsilon^{\sigma\lambda\alpha\beta}}{2q\cdot n}q_{\alpha}n_{\beta}\mathcal{D}_{\lambda}
		+\tilde{F}^{\sigma\lambda}q_{\lambda}\delta'(q^2)
		\bigg]f_{\rm L}
		\\\nonumber
		&=&-4\pi \hbar c \bigg[\delta(q^2)\Big(q_{[\nu}\mathcal{D}_{\mu]}+\frac{q^2}{q\cdot n}n_{[\mu}\mathcal{D}_{\nu]}+\frac{q_{[\mu}n_{\nu]}}{q\cdot n}q\cdot\mathcal{D}\Big)
		+\big(2q^{\rho}F_{\rho[\mu}q_{\nu]}+q^2F_{\mu\nu}\big)\delta'(q^2)
		\bigg]f_{\rm L}
		\\
		&=&-2\pi \hbar c \Big[\delta(q^2)\big(2F_{\nu\mu}+2q_{[\nu}\mathcal{D}_{\mu]}\big)+4q^{\rho}F_{\rho[\mu}q_{\nu]}\delta'(q^2)\Big]f_{\rm L}+O(\hbar^2),
		\label{epsilon_q_L}
	\end{eqnarray}
	where we used $q^2 \delta(q^2)=0$ and $q^2 \delta'(q^2) = -\delta(q^2)$ in the third line.
	It is thus clear that Equation~(\ref{WF_L_full}) satisfies Equation~(\ref{m_eq3}).

\section{Conservative equations}
\label{app_conservative}
\setcounter{equation}{0}
We derive the conservative equation for the energy-momentum tensor in the curve spacetime. For simplicity, we consider the case without electromagnetic fields. We find
\begin{eqnarray}\nonumber
2\int_q D\cdot \mathcal{L}^< 
&=&2\int_q \left(\partial_{\mu}+\Gamma^{\rho}_{\rho\mu}-\Gamma^{\lambda}_{\mu\rho}q^{\rho}\partial_{q\lambda}\right)\mathcal{L}^{<\mu}
\\\nonumber
&=&2\partial_{\mu}\int_q \mathcal{L}^{<\mu}
-2\int_q \left(\Gamma^{\rho}_{\mu\rho}\mathcal{L}^{<\mu}-\Gamma^{\rho}_{\rho\mu}\mathcal{L}^{<\mu}-\Gamma^{\rho}_{\mu\rho}\mathcal{L}^{<\mu}\right)
\\\nonumber
&=&2\left(\partial_{\mu}+\Gamma^{\rho}_{\rho\mu}\right)\int\frac{{\rm d}^4q}{(2\pi)^4}\mathcal{L}^{<\mu}
\\
&=&\nabla_{\mu}J^{\mu},
\end{eqnarray}
where we used integration by parts and dropped the surface terms, and employed the relation
\begin{eqnarray}
\partial_{\mu}\sqrt{-g}=\frac{1}{2}\sqrt{-g}g^{\alpha\beta}\partial_{\mu}g_{\alpha\beta}=\sqrt{-g}\Gamma^{\rho}_{\mu\rho}\,.
\end{eqnarray}
Note that $\nabla_{\mu}\int {\rm d}^4q\neq 0$. Recall that the master equation with the collisions reads
$D\cdot\mathcal{L}^<=\check{\mathcal{C}}[f]$,
where $\check{\mathcal{C}}[f]$ is defined in Equation~(\ref{C_check}). 
We hence obtain the conservative equation for the particle-number current,
\begin{eqnarray}
\nabla\cdot J=2\int_q \check{\mathcal{C}}[f]\,.
\end{eqnarray}

For the energy-momentum tensor, on the other hand, we start with
\begin{eqnarray}\nonumber
2\int_q q^{\{\nu}D_{\mu}\mathcal{L}^{<\mu\}}&=&
2\int_q q^{\{\nu}\left(\partial_{\mu}\mathcal{L}^{<\mu\}}+\Gamma^{\mu\}}_{\mu\rho}\mathcal{L}^{<\rho}-\Gamma^{\lambda}_{\mu\rho}q^{\rho}\partial_{q\lambda}\mathcal{L}^{<\mu\}}\right)
\\\nonumber
&=&2\int\frac{{\rm d}^4q}{(2\pi)^4}\Big[\big(\partial_{\mu}q^{\{\nu}+\Gamma^{\{\nu}_{\mu\rho}q^{\rho}\big)\sqrt{-g}\mathcal{L}^{<\mu\}}
+q^{\{\nu}\Gamma^{\mu\}}_{\mu\rho}\sqrt{-g}\mathcal{L}^{<\rho}\Big]
\\\nonumber
&=&\partial_{\mu}T^{\mu\nu}+\Gamma^{\mu}_{\mu\rho}T^{\rho\nu}+\Gamma^{\nu}_{\mu\rho}T^{\mu\rho}
\\
&=&\nabla_{\mu}T^{\mu\nu},
\end{eqnarray}
where we utilized a similar trick to that in the case for the particle-number current. Following the trick in \cite{Gorbar:2017toh,Hidaka:2018ekt}, we find
\begin{eqnarray}\nonumber\label{div_T_1}
\nabla_{\mu}T^{\mu\nu}&=&\int_q \Big(2q^{\nu}D\cdot \mathcal{L}^<+q\cdot D \mathcal{L}^{<\nu}-q^{\nu}D\cdot \mathcal{L}^<\Big)
\\
&=&\int_q \Big(2q^{\nu}D\cdot \mathcal{L}^<-\frac{1}{2}\epsilon^{\nu\kappa\sigma\rho}\epsilon_{\mu\lambda\sigma\rho}q^{\lambda}D_{\kappa}\mathcal{L}^{<\mu}\Big)\,.
\end{eqnarray}
By using the second line of Equation~(\ref{epsilon_q_L}) without electromagnetic fields, one obtains
\begin{eqnarray}
\epsilon_{\mu\lambda\sigma\rho}q^{\lambda}D_{\kappa}\mathcal{L}^{<\mu}
=2\pi \hbar c \delta(q^2)D_{\kappa}\Big(q_{[\rho}\mathcal{D}_{\sigma]}f_{\rm L}+\frac{q_{[\sigma}n_{\rho]}}{q\cdot n}q\cdot \mathcal{D}f_{\rm L}\Big)
=2\pi \hbar c \delta(q^2)D_{\kappa}\Big(q_{[\rho}\mathcal{D}_{\sigma]}f_{\rm L}\Big)+O(\hbar^2)\,,
\end{eqnarray}
and consequently, 
\begin{eqnarray}
-\frac{1}{2}\epsilon^{\nu\kappa\sigma\rho}\epsilon_{\mu\lambda\sigma\rho}q^{\lambda}D_{\kappa}\mathcal{L}^{<\mu}
\approx-\pi \hbar c \delta(q^2)\epsilon^{\nu\kappa\sigma\rho}q_{\rho}D_{\kappa}\mathcal{D}_{\sigma}f_{\rm L}
=\frac{\pi}{2}\hbar c \delta(q^2)q_{\rho}\epsilon^{\nu\rho\kappa\sigma}\Big(q^{\alpha}R^{\beta}_{\,\alpha\kappa\sigma}\partial_{q\beta}f_{\rm L}
+2D_{\kappa}\mathcal{C}_{\sigma}[f_{\rm L}] \Big)\,.
\end{eqnarray} 
Nonetheless, the term involving the Riemann curvature tensor will vanish when integrating over momentum as
\begin{eqnarray}
\int_q \delta(q^2)q_{\rho}\epsilon^{\nu\rho\kappa\sigma}q^{\alpha}R^{\beta}_{\,\alpha\kappa\sigma}\partial_{q\beta}f_{\rm L}=\int_q \delta(q^2)q^{\alpha}\epsilon^{\nu\rho\kappa\sigma}R_{\alpha\rho\kappa\sigma}f_{\rm L}=0\,,
\end{eqnarray}
where we used integration by parts and dropped the surface terms again, and we implemented the following properties for Riemann curvature tensor:
\begin{eqnarray}
R_{\alpha\rho\kappa\sigma}=-R_{\rho\alpha\kappa\sigma}=-R_{\alpha\rho\sigma\kappa},
\end{eqnarray}
and the Bianchi identity
\begin{eqnarray}
R_{\alpha\rho\kappa\sigma}+R_{\alpha\kappa\sigma\rho}+R_{\alpha\sigma\rho\kappa}=0.
\end{eqnarray}
Similar results can be found for right-handed fermions by flipping the sign of the $O(\hbar)$ term. 
Therefore, from Equation~(\ref{div_T_1}), we derive
\begin{eqnarray}
\nabla_{\mu}T^{\mu\nu}
=2\int_q \bigg(q^{\nu}\check{\mathcal{C}}[f_{\rm L}]
+\frac{\pi}{2}\hbar c \delta(q^2)\epsilon^{\nu\mu\alpha\beta}q_{\alpha}D_{\beta}\mathcal{C}_{\mu}[f_{\rm L}]\bigg)
\end{eqnarray}
as the conservative equation for the energy-momentum tensor.

One may sometimes write the conservative equation in an alternative form.
By using the Schouten identity, which states that an antisymmetric tensor of rank 5 vanishes in 4 spacetime dimensions, i.e.,
\begin{eqnarray}
0 = 5 \delta_{\sigma}^{[\nu}\epsilon^{\mu \rho \alpha \beta]} \equiv  \delta_{\sigma}^{\nu} \epsilon^{\mu \rho \alpha \beta}
+ \delta_{\sigma}^{\mu} \epsilon^{\rho \alpha \beta \nu} + \delta_{\sigma}^{\rho} \epsilon^{\alpha \beta \nu \mu}
+ \delta_{\sigma}^{\alpha} \epsilon^{\beta \nu \mu \rho} + \delta_{\sigma}^{\beta} \epsilon^{\nu \mu \rho \alpha},
\end{eqnarray}
one finds
\begin{eqnarray}
q^{\nu}\check{\mathcal{C}}[f_{\rm L}]=2\pi\delta(q^2)\bigg(q^{\nu} (q\cdot\mathcal{C}) +\hbar c \frac{\epsilon^{\nu\mu\alpha\beta}}{2q\cdot n}\Big[q_{\alpha}n_{\beta}(q\cdot\Sigma^<+q\cdot\Sigma^>)D_{\mu}+q_{\alpha}q\cdot n(\Sigma^<_{\mu}+\Sigma^>_{\mu})D_{\beta}\Big]f_{\rm L}\bigg)+O(\hbar^2)\,,
\end{eqnarray}
which leads to
\begin{eqnarray}\nonumber
\nabla_{\mu}T^{\mu\nu}
&=&4\pi \int_q \delta(q^2)\bigg(q^{\nu} (q\cdot\mathcal{C})
+\frac{1}{4}\hbar c \epsilon^{\nu\mu\alpha\beta}q_{\alpha}\bigg[\frac{2n_{\beta}}{q\cdot n}(q\cdot\Sigma^<+q\cdot\Sigma^>)D_{\mu}f_{\rm L}
+(\Sigma^<_{\mu}+\Sigma^>_{\mu})D_{\beta}f_{\rm L}
\\
&&+\Big((1-f_{\rm L})D_{\beta}\Sigma^<_{\mu}-f_{\rm L}D_{\beta}\Sigma^{>}_{\mu}\Big)
\bigg]\bigg)\,.
\end{eqnarray}

\section{Collision Terms in the Four-Fermi Theory}
\label{app_four-Fermi_theory}
\setcounter{equation}{0}
We can write down an explicit form of the collision terms in $\check{\mathcal{C}}$ for the neutrino absorption process in Equation~(\ref{inelastic_s}) as
\begin{eqnarray}\nonumber\label{S_Sigma}
\mathcal{L}^{(\nu)\lessgtr}_q\cdot\Sigma^{\gtrless}
&&=\frac{{G}_{\rm F}^2}{2}\int_{k,q',k'}\bigg[\text{Tr}\Big(\grave{S}^{(\nu)\lessgtr}_{{\rm L}q}\gamma_{\mu}\grave{S}^{({\rm e})\gtrless}_{{\rm L}q'}\gamma_{\rho}\Big)
\text{Tr}\Big(W^{\lessgtr}_{k}\gamma^{\mu}(g_{\rm V}-g_{\rm A}\gamma^5)W'^{\gtrless}_{k'}\gamma^{\rho}(g_{\rm V}-g_{\rm A}\gamma^5)\Big)\bigg]
\\\nonumber
&&=8{G}_{\rm F}^2\int_{k,q',k'}\bigg[(g_{\rm V}^2+g_{\rm A}^2)\Big(k\cdot \mathcal{L}^{(\nu)\lessgtr}_{q}k'\cdot \mathcal{L}^{({\rm e})\gtrless}_{q'}+k'\cdot \mathcal{L}^{(\nu)\lessgtr}_{q}k\cdot \mathcal{L}^{({\rm e})\gtrless}_{q'}\Big)
-(g_{\rm V}^2-g_{\rm A}^2)M_{\rm n}M_{\rm p}\mathcal{L}^{(\nu)\lessgtr}_{q}\cdot \mathcal{L}^{({\rm e})\gtrless}_{q'}
\\
&&
\quad+2g_{\rm V}g_{\rm A}\big(\mathcal{L}^{(\nu)\lessgtr}_{q}\cdot k\mathcal{L}^{({\rm e})\gtrless}_{q'}\cdot k'-\mathcal{L}^{(\nu)\lessgtr}_{q}\cdot k'\mathcal{L}^{({\rm e})\gtrless}_{q'}\cdot k\big)
\bigg](2\pi)^2\delta(k^2-M_{\rm n}^2)\delta(k'^2-M_{\rm p}^2)f^{({\rm n})\lessgtr}_{k}f^{({\rm p})\gtrless}_{k'}\,.
\end{eqnarray}
Here $\grave{S}^{(\nu/{\rm e})\lessgtr}_{{\rm L}q}=P_{\rm L} \gamma^{\mu}\mathcal{L}^{(\nu/{\rm e})\lessgtr}_{q\mu}$ are the lesser/greater propagators for left-handed neutrinos and electrons with $\mathcal{L}^{(\nu/{\rm e})\lessgtr}_{q\mu}$ defined in Equations (\ref{L_nu}) and (\ref{L_e}), and $W^{\lessgtr}$ and $W'^{\lessgtr}$ are the lesser/greater propagators for nucleons, which we assumed to take the simple form of the free Dirac fermions,
\begin{eqnarray}
W^{\lessgtr}(k)=2\pi\delta(k^2-M_{\rm N}^2)(\slashed{k}-M_{\rm N})f^{({\rm N})\lessgtr}_{k},
\end{eqnarray} 
and similarly for $W'^{\lessgtr}$, and we ignored the antinucleons. Here we also defined
\begin{eqnarray}
\int_{k,q',k'}=\int\frac{{\rm d}^4k{\rm d}^4q'{\rm d}^4k'}{(2\pi)^8}(\sqrt{-g})^3 \delta^{(4)}(q+k-q'-k')\,.
\end{eqnarray}
The self-energies then read
\begin{eqnarray}
\Sigma^{\gtrless}_{\mu}
&=& 8{G}_{\rm F}^2\int_{k,q',k'}\left(g_{+}^2k_{\mu}k'\cdot \mathcal{L}^{({\rm e})\gtrless}_{q'}
+g_{-}^2k'_{\mu}k\cdot \mathcal{L}^{({\rm e})\gtrless}_{q'}-g_{+}g_{-}M_{\rm n}M_{\rm p}\mathcal{L}^{({\rm e})\gtrless}_{q'\mu} \right)
(2\pi)^2\delta(k^2-M_{\rm n}^2)\delta(k'^2-M_{\rm p}^2)f^{({\rm n})\lessgtr}_{k}f^{({\rm p})\gtrless}_{k'}
\nonumber \\
&=& 8{G}_{\rm F}^2\int_p\mathcal{L}^{({\rm e})\gtrless\nu}_{q-p}\int_k \left(g_{+}^2k_{\mu}k'_{\nu} 
+g_{-}^2k'_{\mu}k_{\nu}-g_{+}g_{-}M_{\rm n}M_{\rm p}\eta_{\mu\nu} \right)
(2\pi)^2\delta(k^2-M_{\rm n}^2)\delta(k'^2-M_{\rm p}^2)f^{({\rm n})\lessgtr}_{k}f^{({\rm p})\gtrless}_{k'} \bigg|_{k'=p+k}\,.
\nonumber \\
\label{self_E}
\end{eqnarray} 

The self-energy for the elastic scattering in Equation~(\ref{elastic_s}) can be obtained by the following replacements in Equation~(\ref{self_E}): ${\rm n}\rightarrow {\rm N}$, ${\rm p}\rightarrow {\rm N}$, $g_{{\rm V,A}}\rightarrow c_{{\rm V,A}}/2$, and $\mathcal{L}^{({\rm e})\gtrless}_{q'\mu}\rightarrow \mathcal{L}^{(\nu)\gtrless}_{q'\mu}$.

\section{Self-Energies in Equilibrium}
\label{app_self_energies}
\setcounter{equation}{0}
We describe the details of the calculations of the self-energies $\bar \Sigma^{(0)\gtrless}_{\mu}$, $\bar \Sigma^{(\omega)\gtrless}_{\mu}$, and $\bar \Sigma^{(B)\gtrless}_{\mu}$ under certain approximations. We first consider the nonrelativistic limit for nucleons, where $k^{\mu}\approx M_{\rm n} u^{\mu}+k_{\perp}^{\mu}$. Here we introduced $V_0\equiv V\cdot u$ and $V^{\mu}_{\perp}\equiv V^{\mu}-(V\cdot u)u^{\mu}$ for an arbitrary four-vector $V^{\mu}$. In addition, we will use $|{\bm V}|$ to represent the norm of $V^{\mu}_{\perp}$. 
Then, we can approximate the following factors appearing in Equations (\ref{Sigma0eq})--(\ref{Sigma_B_eq}) as
\begin{eqnarray}\nonumber
&& \! \! \! \! \! \! \! \! \!
\Big(g_{+}^2k_{\mu}k'_{\nu}+g_{-}^2 k'_{\mu}k_{\nu}-g_{+}g_{-}M_{\rm n}M_{\rm p}\eta_{\mu\nu} \Big) q'^{\nu}
\\
&\approx& \Big[\big(g_+^2+g_-^2\big)M_{\rm n}^2-g_+g_-M_{\rm n}M_{\rm p}\Big](q_0-p_0)u_{\mu}-g_{+}g_{-}M_{\rm n}M_{\rm p}(q-p)_{\perp\mu}\,, \\
\nonumber
&& \! \! \! \! \! \! \! \! \!
g_{+}^2k_{\mu}\big[ (\omega \cdot k')(q' \cdot u) - (\omega \cdot q')(k' \cdot u) \big]
+g_{-}^2 k'_{\mu}\big[(\omega \cdot k)(q' \cdot u) - (\omega \cdot q')(k\cdot u) \big]
-g_{+}g_{-}M_{\rm n}M_{\rm p}\big[\omega_{\mu} (q' \cdot u) - u_{\mu} (q' \cdot\omega) \big]
\\
&\approx&-\Big[\big(g_+^2+g_-^2\big)M_{\rm n}^2-g_+g_-M_{\rm n}M_{\rm p}\Big]\omega\cdot (q-p)u_{\mu}-g_{+}g_{-}M_{\rm n}M_{\rm p}(q_0-p_0)\omega_{\mu}\,, \\
\nonumber
&& \! \! \! \! \! \! \! \! \!
g_{+}^2k_{\mu}\big[(B \cdot k')(q'\cdot u) - (B \cdot q')(k'\cdot u) \big]+g_{-}^2 k'_{\mu}\big[(B \cdot k)(q' \cdot u)-(B \cdot q')(k\cdot u)\big]
-g_{+}g_{-}M_{\rm n}M_{\rm p}\big[B_{\mu} (q'\cdot u) - u_{\mu} (q'\cdot B) \big]
\\
&\approx&-\Big[\big(g_+^2+g_-^2\big)M_{\rm n}^2-g_+g_-M_{\rm n}M_{\rm p}\Big]B\cdot (q-p)u_{\mu}-g_{+}g_{-}M_{\rm n}M_{\rm p}(q_0-p_0)B_{\mu}\,,
\end{eqnarray}
where we used $k'=p+k$ and $q'=q-p$.
Following \cite{Reddy:1997yr}, we may drop all the last terms above owing to the small prefactors proportional to $g_+g_-=g_{\rm V}^2-g_{\rm A}^2$, while we still keep the $g_+g_-$ terms proportional to $u^{\mu}$. Also, we will ignore the difference of neutron and proton masses and set $M_{\rm n} \approx M_{\rm p} \approx M$. Then, one may combine $\bar \Sigma^{(0)\gtrless}_{\mu}$ and $\bar \Sigma^{(B)\gtrless}_{\mu}$ as
\begin{eqnarray}\label{Sigma0+Sigma_B}\nonumber
\bar \Sigma^{(0)\gtrless}_{\mu}+\hbar \bar \Sigma^{(B)\gtrless}_{\mu}&\approx & 
8 u_{\mu}\big(g_+^2+g_-^2-g_+ g_-\big)M^2{G}_{\rm F}^2\int_p\int_k(q_0-p_0)(2\pi)^3\delta\left((q-p)^2+\frac{\hbar B\cdot(q-p)}{q_0-p_0}\right)\delta(k^2-M^2)
\\
&&\times \delta((k+p)^2-M^2)
f^{({\rm n})\lessgtr}_{0,k} f^{({\rm p})\gtrless}_{0,p+k} f^{({\rm e})\gtrless}_{0,q-p}\,.
\end{eqnarray}

We next exploit the isoenergetic approximation in \cite{Bruenn:1985en} by taking $p_0\rightarrow 0$, which in fact can be realized under the nonrelativistic approximation for nucleons. In the nonrelativistic limit, one can rewrite the following delta function as
\begin{eqnarray}\label{iso_E_approx}
 \delta((k+p)^2-M^2)=\delta(p^2+2p\cdot k)
\approx\frac{1}{2M}\delta\Big(p_0-\frac{|{\bm p}|^2}{2M}-\frac{|{\bm p}||{\bm k}|\cos\theta_{pk}}{M}\Big) 
\end{eqnarray}
by dropping higher-order terms suppressed by $1/M$, where $\theta_{pk}$ denotes the angle between ${\bm k}$ and ${\bm p}$. When further neglecting the $O(|{\bm p}|/M)$ and $O(|{\bm k}|/M)$ terms, the delta function can be approximated as $ \delta((k+p)^2-M^2)\approx \delta(p_0)/(2M)$, which explicitly yields $p_0\rightarrow 0$. Nonetheless, in order to include the quantum corrections from magnetic fields, we should retain at least the $O(|{\bm p}|/M)$ terms in Equation~(\ref{iso_E_approx}). In contrast, we may omit the $|{\bm k}|\cos\theta_{pk}/M$ term by symmetry when assuming that $f^{({\rm p})\gtrless}_{p+k}$ only depends on $p_0+k_0$ given that the nucleons are near thermal equilibrium. Physically, when $M\rightarrow \infty$, the momentum transfer characterized by $|{\bm p}|$ is suppressed, while the magnetic field affects the momentum of the outgoing electron in scattering. Based on momentum conservation, it is hence necessary to include at least $|{\bm p}|/M$ terms for preserving the magnetic field contribution albeit the suppression in the nonrelativistic limit. We will accordingly apply $ \delta((k+p)^2-M^2)=\delta(p^2+2p\cdot k)\approx (2M)^{-1}\delta\Big(p_0-|{\bm p}|^2/(2M)\Big)$ as a ``quasi-isoenergetic" approximation.
On the other hand, for an arbitrary integrand $G(p,k)$, one can write the integral as
\begin{eqnarray}\nonumber
&&\int_p\int_k\delta\left((q-p)^2+\frac{\hbar B\cdot(q-p)}{q_0-p_0}\right)\delta(k^2-M^2)G(p,k)
\\
&&\approx\int\frac{{\rm d}p_0 {\rm d}|{\bm p}|{\rm d}(\cos\theta_{pq})}{(2\pi)^2}\frac{|{\bm p}|}{2|{\bm q}|}\left(1-\frac{\hbar B_{\rm L}}{2|{\bm q}|(q_0-p_0)}\right)\delta(\cos\theta_{pq}-\cos\theta_{B})
\int\frac{{\rm d}^3{\bm k}}{(2\pi)^32E_k}G(p,k)\big|_{k_0=E_k}\,,
\end{eqnarray}
where we decomposed the magnetic field as Equation~(\ref{B_decomp}) and
\begin{eqnarray}
\cos\theta_{B} \equiv \frac{1}{2|{\bm p}||{\bm q}|}\left[\left(1-\frac{\hbar B_{\rm L}}{2|{\bm q}|(q_0-p_0)}\right)\big(2q_0p_0-p^2\big)+\frac{\hbar B_{\rm L}|{\bm q}|}{q_0-p_0}\right]\,.
\end{eqnarray}
Here we applied 
\begin{eqnarray}\nonumber\label{disp_B_field}
\delta\left((q-p)^2+\frac{\hbar B\cdot(q-p)}{q_0-p_0}\right)&=&\delta\left(\Big(2|{\bm q}|+\frac{\hbar B_{\rm L}}{q_0-p_0}\Big)|{\bm p}|\cos\theta_{pq}+p^2-2q_0p_0-\hbar \frac{B_{\rm L}|{\bm q}|+B_{\rm T}\cdot p}{q_0-p_0}\right)
\\
&\approx& \frac{1}{2|{\bm p}||{\bm q}|}\left(1-\frac{\hbar B_{\rm L}}{2|{\bm q}|(q_0-p_0)}\right)\delta(\cos\theta_{pq}-\cos\theta_{B})\,,
\end{eqnarray}
where $B_{\rm T}\cdot p=-B_{\rm T}|{\bm p}|\sin\theta_{pq}\cos\phi_{pB}$ is also dropped by assuming $|{\bm q}|\gg |{\bm p}|$ in the derivation.%
\footnote{Although this assumption is not rigorously justified, the contribution from $B_{\rm T}$ will eventually be irrelevant regardless of this assumption, since the term involving the magnetic field after the integral must be proportional to $q\cdot B$ from the symmetry, which can only incorporate the contribution from $B_{\rm L}$.} In the following, we will set $q_0 = |{\bm q}|$ based on the on-shell condition for neutrinos.

Based on the quasi-isoenergetic approximation, further assuming that $f^{({\rm n})\lessgtr}_{k}$ and $f^{({\rm p})\gtrless}_{p+k}$ only depend on $k_0$ and $p_0+k_0$, we find
\begin{eqnarray}
\bar \Sigma^{(0)\gtrless}_{\mu}+\hbar \bar \Sigma^{(B)\gtrless}_{\mu}\approx  
4\pi u_{\mu}\big(g_{\rm V}^2+3g_{\rm A}^2\big){G}_{\rm F}^2 f^{({\rm e})\gtrless}_{0,q}\int^{p_{\text{max}}}_{p_{\text{min}}}\frac{{\rm d}|{\bm p}|}{(2\pi)^2}\frac{|{\bm p}|}{2|{\bm q}|}\left(|{\bm q}|-\frac{|{\bm p}|^2}{2M}-\frac{\hbar B_{\rm L}}{2|{\bm q}|}\right)
\int\frac{{\rm d}^3{\bm k}}{(2\pi)^3}
f^{({\rm n})\lessgtr}_{0,k} f^{({\rm p})\gtrless}_{0,k}\,,
\end{eqnarray}
where we rewrote $g_{\pm}$ in terms of $g_{\rm V, A}$. Here $p_{\text{max}}$ and $p_{\text{min}}$ are determined by the dispersion relation in Equation~(\ref{disp_B_field}) as
\begin{eqnarray}
p_{\text{max}}=|{\bm q}|\left(2-\frac{2|{\bm q}|}{M}+\frac{\hbar B_{\rm L}}{2|{\bm q}|^2}\right),\qquad p_{\text{min}}=\frac{\hbar B_{\rm L}}{2|{\bm q}|}\,.
\end{eqnarray}
Using the following relations for the nucleon Fermi-Dirac distribution, 
\begin{eqnarray}
f_{0,k}^{({\rm n})<} f_{0,k}^{({\rm p})>}=\frac{f^{({\rm n})}_{0,k}-f^{({\rm p})}_{0,k}}{1-{\rm e}^{\beta(\mu_{{\rm p}}-\mu_{\rm n})}},\qquad 
f_{0,k}^{({\rm n})>} f_{0,k}^{({\rm p})<}=\frac{f^{({\rm p})}_{0,k}-f^{({\rm n})}_{0,k}}{1-{\rm e}^{\beta(\mu_{\rm n}-\mu_{{\rm p}})}}\,,
\end{eqnarray} 
one finds Equation (\ref{Sigma_mu_0B}).

Additionally, we also have the collision terms associated with the fluid vorticity,
\begin{eqnarray}
	\hbar q\cdot \bar \Sigma^{(\omega)\gtrless}\approx  
	\pm 2\pi \hbar \big(g_{\rm V}^2+3g_{\rm A}^2\big){G}_{\rm F}^2 \beta(q\cdot \omega)  f^{({\rm e})}_{0,q}(1-f^{({\rm e})}_{0,q})\int^{2|{\bm q}|}_{0}\frac{{\rm d}|{\bm p}|}{(2\pi)^2}\frac{|{\bm p}|}{2}
	\int\frac{{\rm d}^3{\bm k}}{(2\pi)^3}
	f^{({\rm n})\lessgtr}_{0,k} f^{({\rm p})\gtrless}_{0,k}\,,
\end{eqnarray}
and 
\begin{eqnarray}\nonumber
	\hbar S^{\mu\nu}_{(\xi)}D_{\mu} \bar \Sigma^{(0)\gtrless}_{\nu}&=& 4\pi \hbar \big(g_{\rm V}^2+3g_{\rm A}^2\big){G}_{\rm F}^2 \frac{\epsilon^{\mu\nu\alpha\beta}q_{\alpha}\xi_{\beta}}{2q\cdot\xi}
	(\nabla_{\mu}u_{\nu}) f^{({\rm e})\gtrless}_{0,q} \int^{2|{\bm q}|}_{0}\frac{{\rm d}|{\bm p}|}{(2\pi)^2}\frac{|{\bm p}|}{2}
	\int\frac{{\rm d}^3{\bm k}}{(2\pi)^3}
	f^{({\rm n})\lessgtr}_{0,k} f^{({\rm p})\gtrless}_{0,k}
	\\\nonumber
	&=&4\pi \hbar \big(g_{\rm V}^2+3g_{\rm A}^2\big){G}_{\rm F}^2\frac{1}{q\cdot\xi} \big[(q\cdot \omega)(u\cdot\xi)-(q \cdot u)(\omega\cdot\xi)\big] f^{({\rm e})\gtrless}_{0,q}\int^{2|{\bm q}|}_{0}\frac{{\rm d}|{\bm p}|}{(2\pi)^2}\frac{|{\bm p}|}{2}
	\int\frac{{\rm d}^3{\bm k}}{(2\pi)^3}
	f^{({\rm n})\lessgtr}_{0,k} f^{({\rm p})\gtrless}_{0,k}
	\\
	&\approx&4\pi \hbar \big(g_{\rm V}^2+3g_{\rm A}^2\big){G}_{\rm F}^2 \frac{q\cdot \omega}{|{\bm q}|} f^{({\rm e})\gtrless}_{0,q}\int^{2|{\bm q}|}_{0}\frac{{\rm d}|{\bm p}|}{(2\pi)^2}\frac{|{\bm p}|}{2}
	\int\frac{{\rm d}^3{\bm k}}{(2\pi)^3}
	f^{({\rm n})\lessgtr}_{0,k} f^{({\rm p})\gtrless}_{0,k}\,,
\end{eqnarray}
where we used $\xi^{\mu}\approx u^{\mu}$ and dropped the higher-order terms in $|{\bm v}|$. We hence arrive at Equation (\ref{Sigma_mu_omega}).

\section{Conservative Equation in Equilibrium}
\label{app_conserv_eq}
\setcounter{equation}{0}
Let us consider the conservative equation for the energy-momentum tensor in Equation~(\ref{cons_Tmunu}) when the matter sector is in thermal equilibrium.
When taking $n^{\mu}=\xi^{\mu}\approx u^{\mu}$, the $\hbar$ term in $\check{\mathcal{C}}[f_{\rm L}^{(\nu)}]$ vanishes since now $\Sigma^{\lessgtr}_{\mu}\sim u_{\mu}$. 
We thus focus on the second term on the right-hand side of Equation~(\ref{cons_Tmunu}).
By using
\begin{eqnarray}\label{nabla_rel_1}
	\epsilon^{\nu\mu\alpha\beta}q_{\alpha}\nabla_{\beta}\left[ u_{\mu}(q\cdot u)^2 \right]=2(q\cdot u) \big[2(q\cdot u)^2 \omega^{\nu} - (q\cdot u)(q\cdot \omega) u^{\nu}
	+ (q\cdot \omega) q_{\perp}^{\nu} \big]
\end{eqnarray}
and 
\begin{eqnarray}\label{nabla_rel_2}
	\epsilon^{\nu\mu\alpha\beta}q_{\alpha}u_{\mu}\nabla_{\beta} f_{0,q}^{({\rm e})}=-\beta f_{0,q}^{({\rm e})}(1-f_{0,q}^{({\rm e})})
	[(q\cdot u)^2 \omega^{\nu} + (q\cdot \omega)q_{\perp}^{\nu}]
\end{eqnarray}
under the on-shell condition $q^2=0$, we have
\begin{eqnarray}\nonumber
	\epsilon^{\nu\mu\alpha\beta}q_{\alpha}D_{\beta} \bar \Sigma^{(0)>}_{\mu}&=&
	\bigg(|{\bm q}|^2 \big(4 + \beta |{\bm q}| f_{0,q}^{({\rm e})} \big) \omega^{\nu}
	+ q\cdot \omega \Big[\big(2+\beta |{\bm q}| f_{0,q}^{({\rm e})} \big)q_{\perp}^{\nu} - 2|{\bm q}|u^{\nu} \Big] \bigg)
	\frac{q\cdot \bar \Sigma^{(0)>}}{|{\bm q}|^2}\,,
	\\
	\epsilon^{\nu\mu\alpha\beta}q_{\alpha}D_{\beta} \bar \Sigma^{(0)<}_{\mu}&=&
	\bigg(|{\bm q}|^2 \big(4 - \beta |{\bm q}| (1-f_{0,q}^{({\rm e})}) \big) \omega^{\nu}
	+ q\cdot \omega \Big[\big(2-\beta |{\bm q}| (1-f_{0,q}^{({\rm e})}) \big)q_{\perp}^{\nu} - 2|{\bm q}|u^{\nu} \Big] \bigg)
	\frac{q\cdot \bar \Sigma^{(0)<}}{|{\bm q}|^2}\,,
\end{eqnarray}
from which we derive%
\footnote{Here the (four-)momentum derivatives in the horizontal lifts acting on $\bar{\Sigma}^{(0)\lessgtr}_{\mu}$ do not contribute, as is consistent with the physical expectation that the $\hbar$ corrections will depend on either $\omega_{\mu}$ or $B_{\mu}$ in local thermal equilibrium. Technically, we find that $\bar{\Sigma}^{(0)\lessgtr}_{\mu}=u_{\mu}\mathcal{F}^{\lessgtr}(q\cdot u)$ with $\mathcal{F}^{\lessgtr}(q\cdot u)$ being functions of $q_0$ alone, and combined with the relations $\Gamma^{0}_{ab}=\Gamma^{c}_{0b}=\Gamma^{c}_{a0}=0$ in the inertial frame, the (four-)momentum derivatives lead to vanishing results after contracting with the Christoffel symbols. A similar argument is applied to $u_{\rho}\nabla_{\mu}q^{\rho}=0$ in Equations~(\ref{nabla_rel_1}) and (\ref{nabla_rel_2}).}
\begin{eqnarray}\nonumber
	&&\hbar \epsilon^{\nu\mu\alpha\beta}q_{\alpha}D_{\beta}\mathcal{C}_{\mu}[f_{\rm L}^{(\nu)}]
	\\\nonumber
	&&\approx -\hbar \epsilon^{\nu\mu\alpha\beta}q_{\alpha}\Big(\bar \Sigma^{(0)>}_{\mu}+\bar \Sigma^{(0)<}_{\mu}\Big)D_{\beta}f_{{\rm L}q}^{(\nu)}
	+ \hbar(1-f_{{\rm L}q}^{(\nu)})\bigg(|{\bm q}|^2 \big(4 - \beta |{\bm q}| (1-f_{0,q}^{({\rm e})}) \big) \omega^{\nu}
	+ q\cdot \omega \Big[\big(2-\beta |{\bm q}| (1-f_{0,q}^{({\rm e})}) \big)q_{\perp}^{\nu} 
	\\
	&& \quad - 2|{\bm q}|u^{\nu} \Big] \bigg)
	\frac{q\cdot \bar \Sigma^{(0)>}}{|{\bm q}|^2}
	- \hbar f_{{\rm L}q}^{(\nu)} \bigg(|{\bm q}|^2 \big(4 + \beta |{\bm q}| f_{0,q}^{({\rm e})} \big) \omega^{\nu}
	+ q\cdot \omega \Big[\big(2+\beta |{\bm q}| f_{0,q}^{({\rm e})} \big)q_{\perp}^{\nu} - 2|{\bm q}|u^{\nu} \Big] \bigg) 
	\frac{q\cdot \bar \Sigma^{(0)<}}{|{\bm q}|^2}\,,
	\label{cons_Tmunu2}
\end{eqnarray}
where the explicit expression of $\bar \Sigma^{(0)\gtrless}_{\mu}$ can be found in Equation~(\ref{Sigma_mu_0B}) by taking $B_{\rm L}=0$.

\newpage
\bibliography{chiral_radiation_hydro_paper_v2.bbl}
\end{document}